\documentclass[epsfig,psfig,aps,twocolumn,prb,showpacs,notitlepage,tightenlines,10pt]{revtex4-2}
\usepackage{graphicx, color}
\usepackage{pbox}
\usepackage{verbatim}
\usepackage{mwe}
\usepackage{amsmath, amsthm, amssymb}
\usepackage{float}
\usepackage{framed}
\usepackage{empheq}
\usepackage{MnSymbol}
\usepackage{bm}
\usepackage{microtype}
\usepackage[colorlinks=true,citecolor=red,urlcolor=blue]{hyperref}
\begin{document}

\title{Scattering of a Dirac particle by a Berry phase domain wall}

\author{Lassaad Mandhour $^{1}$}
\email{lassaad.mandhour@istmt.utm.tn}
\author{Farah Bouhadida $^{1}$}
\author{Fr\'ed\'eric Pi\'{e}chon $^{2}$}
\email{frederic.piechon@universite-paris-saclay.fr}
\affiliation{
$^1$Laboratoire de Physique de la Mati\`ere Condens\'ee, Facult\'e des Sciences de Tunis, Universit\'e Tunis El Manar, Campus Universitaire 1060 Tunis, Tunisia\\
$^2$  Universit\'e Paris-Saclay, CNRS, Laboratoire de Physique des Solides, 91405 Orsay Cedex, France}
\date{\today}
\begin{abstract}
Massless Dirac particles are characterized by an effective pseudospin-momentum locking, which is the origin of the peculiar scattering properties of Dirac particles through potential barriers. This pseudospin-momentum locking also governs the quantum geometric properties (such as the Berry phase and Berry curvature) of Dirac particles. In the present work, we demonstrate that a domain wall separating two regions with distinct quantum geometric properties can serve as an alternative to potential barriers. Specifically, using the three-band
$\alpha-T_3$ model of two-dimensional Dirac particles, we show that a  Berry phase domain wall results in partial reflection and transmission of the Dirac particles, despite the fact that the incident and refracted momenta are identical.

\end{abstract}
\maketitle

\section{Introduction}
 Dirac particles were originally introduced to describe relativistic quantum particles \cite{Dirac1928}. Since then, they have been recognized to emerge as effective quasiparticles in a variety of other physical contexts. For instance, in condensed-matter physics, electronic charge or spin excitations in many materials and heterostructures exhibit behavior akin to two-dimensional (2D) or three-dimensional (3D) massless or massive Dirac quasiparticles \cite{Armitage2018,Lv2021,McClarty2022}. In optics and acoustics, engineered band structures displaying 2D and 3D Dirac cones band crossings have been realized in various photonic and phononic crystals as well as metamaterials \cite{Ozawa2019, Zhang2018,Yang2024}. Cold atomic gases in optical lattices \cite{Cooper2019} and exciton-polaritons in metamaterials have also proven to be highly effective experimental platforms for investigating 2D Dirac particle physics \cite{Ozawa2019,Jaquemin2014}.

In most of these cases, massless Dirac particles are primarily identified by their characteristic “Dirac cones” dispersion relation $E({\bm p}) =c|{\bm p}|$. However, as originally conceived by P. Dirac, a defining property of Dirac particles is that they are pseudospinor waves, $\psi({\bm p})$, which for each momentum is a solution of an effective multiband eigenvalue problem of the form $(c{\bm p} \cdot {\boldsymbol \Lambda})\psi=E({\bm p}) \psi$ where ${\boldsymbol \Lambda}$ represents effective pseudospin matrices
whose specific form depends on the physical system in question \cite{Dirac1928}. Crucially, the pseudospin matrices encode the quantum geometry (such as Berry connection, Berry phase and Berry curvature) of the pseudospinor wavefunction $\psi({\bm p})$, and therefore play a central role in determining the physical properties of the corresponding Dirac particles \cite{Provost1980,Berry1984,Lin2021,Graf2021,Mera2022,Graf2023}.

From this perspective, an intriguing model of 2D Dirac particles that emphasizes the importance of the pseudospin matrices and the associated quantum geometry is the $\alpha-T_3$ model introduced in Ref. \cite{Raoux2014}.
The peculiarity of this model lies in its ability to continuously interpolate, via a parameter $\alpha$, between the honeycomb lattice of graphene ($\alpha=0$), which corresponds to Dirac particles of pseudospin $S=1/2$, and the dice lattice ($\alpha=1$), which corresponds to Dirac particles of pseudospin $S=1$. Notably, while the energy spectrum remains independent of the $\alpha$, the wave functions exhibit an $\alpha$-dependent quantum geometry (Berry phase) \cite{Raoux2014}.

One of the simplest and most common ways to demonstrate the significance of the pseudospin of Dirac particles on their physical properties is by studying their scattering through various kinds of potential barriers. A remarkable phenomenon, known as Klein tunneling, occurs when massless Dirac particles of pseudospin $S$ are perfectly transmitted  at normal incidence \cite{Katsnelson2006,Allain2011}. For other angles of incidence, within the $\alpha-T_3$ model, it has been shown that the transmission probability depends on the pseudospin through the Berry phase of the pseudospinor wavefunctions \cite{Illes2017}. More recently, the role of the pseudospin has been examined for a barrier that combines electric and magnetic potentials \cite{Bouhadida2020}. Beyond potential barriers, other scattering mechanism for massless Dirac particles have been explored, including lattice strain \cite{Yesilyurt, Hung, Zhai, Fujita, Zhai2010, Islam}, line defects \cite{Cheng, Gunlycke,Rodrigues2012,Rodrigues2013,Rodrigues2016,Paez2015}, spatially modulated Fermi velocity \cite{Raoux2010,Concha2010}, interface separating two regions with rotated crystallographic axes in graphene \cite{Romeo2018}, and twist angle disorder in twisted bilayer graphene \cite{Padhi2020}, to name just a few.

In all the aforementioned studies, a key feature is that the scattering of Dirac particles is primarily induced by some external constraint that causes an effective spatial variation of the energy spectrum and/or momentum scattering which in turn leads to an effective pseudospin scattering. In sharp contrast, the present work investigates the scattering of Dirac particles through a domain wall that separates two regions with identical energy spectra but distinct quantum geometries (i.e., distinct Berry phases). Specifically we consider the $\alpha-T_3$ model with a domain wall such that the parameter $\alpha=\alpha_L$ on the left side of the domain wall and $\alpha=\alpha_R$ on the right side  (see Fig. \ref{equiv type2}). A key feature of such a domain wall is that the transmitted (i.e. or refracted) and incident momenta are identical. Despite this, we demonstrate that the Berry phase mismatch at a domain wall leads to a partial reflection-transmission of the Dirac particle.

The paper is organized as follows.   In Sec. \ref{sectionII}, we use a low-energy continuum $\alpha-T_3$  model to compute the transmission probabilities across two classes of effective Berry phase domain walls. In Sec. \ref{sectionIII}, we reconsider the problem using the tight-binding description of the $\alpha-T_3$ model on the Dice lattice.  Within this framework, we examine two kinds of domain walls: a straight (see Fig. \ref{figtype1}) and a zigzag (see Fig. \ref{figtype2}) domain walls. Section \ref{sectionIV} summarizes and concludes the present work.

\section{Continuum description}
\label{sectionII}

   In line with most of the  literature  that studies the scattering of Dirac particles by potential barriers, we use a low-energy continuum  $\alpha-T_3$ model in this section to compute the transmission probabilities across two types of effective Berry phase domain walls. 
We first consider a domain wall without any effective interface pseudopotential.  
Then, in anticipation of the discussion in Section \ref{sectionIII},
we also examine the effect of introducing an effective interface pseudopotential.

\subsection{Dirac particle with a tunable Berry phase}

As will be shown in Sec. \ref{sectionIII}, for the present study, the continuum low-energy effective Hamiltonian for the $\alpha-T_3$ model, expanded around the point $K_\xi=(0,\xi
\frac{4\pi}{3\sqrt{3}})$ in valley $\xi=\pm$, takes a Dirac-like form with a tunable effective pseudospin
\begin{equation}
\begin{array}{ll}
  H_\alpha({\bm q})&=\hbar v_F(\xi q_y S_x^{\alpha}+q_x S_y^{\alpha})\\
  &=\hbar v_F |{\bm q}| \left(
  \begin{array}{ccc}
    0 & -i c_\alpha e^{i \theta}  &0\\
 i c_\alpha e^{-i \theta}&  0 & -i s_\alpha e^{i \theta}   \\
   0 & i s_\alpha e^{-i \theta} &0
  \end{array}
  \right),
  \end{array}
 \label{Hamilton0}
 \end{equation}
 with  $\tan \theta=\frac{\xi q_y}{q_x}$, $c_\alpha=\frac{1}{\sqrt{1+\alpha^2}}$ and $s_\alpha=\frac{\alpha}{\sqrt{1+\alpha^2}}$.
 The pseudo-spin matrices $S_x^{{\alpha}}$ and $S_y^{\alpha}$ interpolate continuously between effective spin-$1/2$ for $\alpha=0$ and spin-1 for $\alpha=1$.
The energy spectrum is composed of two dispersive bands $E_\pm({\bm q})=\pm \hbar v_F\left | {\bm q}\right |$ which form a Dirac cone, and a flat band with energy $E_0=0$. Remarkably, the energy spectrum is independent of the parameter $\alpha$.

By contrast, the corresponding wave functions depend explicitly on the tuning parameter $\alpha$ and write as
\begin{equation}
\psi_{\pm,\theta}(\vec{r})=\frac{1}{\sqrt{2}}\begin{pmatrix}
c_{\alpha} e^{i\theta}\\
\pm i\\
-s_{\alpha} e^{-i\theta}
\end{pmatrix} e^{i\vec{q}\cdot \vec{r}},
\hspace{0.5cm}
\psi_{0,\theta}(\vec{r})=\begin{pmatrix}
s_{\alpha} e^{i\theta}\\
0\\ 
c_{\alpha} e^{-i\theta}
\end{pmatrix} e^{i\vec{q}\cdot \vec{r}}.
\label{psi0}
\end{equation}
As first noted in Ref. \cite{Raoux2014}, this continuous evolution of the wavefunctions with the tuning parameter $\alpha$ translates into an $\alpha$-dependent Berry phase:
$\Phi_{\pm,\xi}(\alpha)=-\xi \pi (1-2s_\alpha^2)$ for the dispersive bands and
$\Phi_{0,\xi}(\alpha)=2\xi \pi (1-2s_\alpha^2)$ for the flat band.

\subsection{Scattering through a Berry phase domain wall}

\begin{figure}[]
       \centering
     \includegraphics[width=0.5\textwidth]{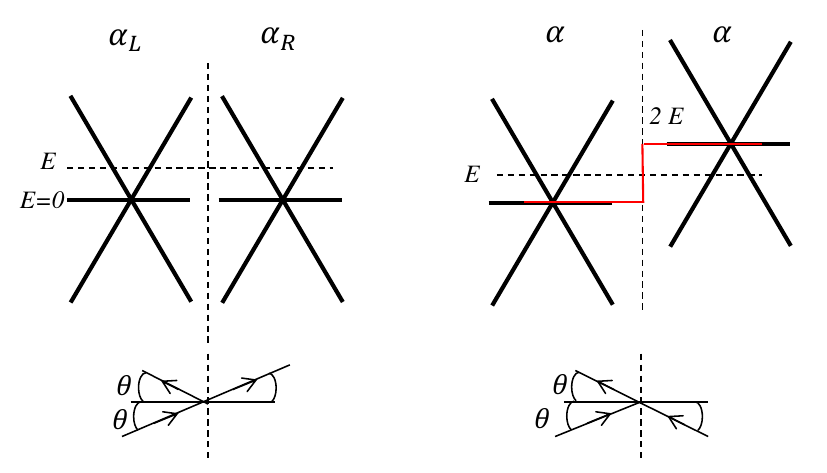}
 \caption{Schematic representation of scattering through the Berry phase domain wall (left) versus through a potential step in the $\alpha-T_3$ model with a step height $V_0$ for an incident energy $E=V_0/2$ (right).}
     \label{equiv type2}
\end{figure}

Many previous works have shown the importance of the $\alpha$-dependent Berry phase on the scattering properties of the low-energy Dirac particle by various kinds of electrostatic and magnetic potential barriers. 
In all these setups, the $\alpha$ parameter was assumed to be spatially constant, so that the pseudospin scattering is proportional to the momentum difference between the transmitted (e.g., refracted) and incident momenta, which is induced by the various potential steps.
 In contrast, we now explore the scattering properties of a Dirac particle through a {\em quantum geometric} domain wall, defined by a step-like variation of the parameter $\alpha(x)$, $\alpha(x)=\alpha_L$ for $x<0$ and $\alpha(x)=\alpha_R$ for $x>0$ as illustrated in Fig. \ref{equiv type2}.
The key peculiarity of this domain wall is that the incident and transmitted momenta are identical which would naively prevent pseudospin scattering. However, as shown below, the $\alpha_L-\alpha_R$ interface effectively induces some pseudospin scattering that is proportional to the Berry phase mismatch $\Phi_s(\alpha_R)-\Phi_s(\alpha_L)$ at the interface.

From now on, to shorten the notations, when there is no ambiguity,
we make the substitution ${\alpha_L,\alpha_R} \rightarrow L,R$.

In the presence of the domain wall, the effective Hamiltonian rewrites
\begin{equation}
 H(x)=
\left\lbrace\begin{array}{ll}
H_{L}(x) & x<0,\\
H_{R}(x) & x>0.
     \end{array}\right.
     \label{simple}
\end{equation}
with
\begin{equation}
H_\alpha(x) =\hbar v_F(-i\partial_x S_y^\alpha +\xi q_y S_x^\alpha),
\end{equation}
for $\alpha=L,R$.

We consider now an incident wave of energy $E>0$ and momentum $(q_x,q_y)=E/\hbar v_F(\cos \theta,\sin \theta)$. Since the transmitted momentum is equal to the incident momentum, and by translational invariance along the $y$ axis, the scattering state writes $\Psi({\bm r}) =e^{i q_y y}\Psi_\theta(x)$ where
$\Psi_\theta(x)=\Psi^L_\theta(x) \theta(-x)+ \Psi^R_{\theta}(x) \theta(x)$. The scattering amplitudes $\Psi^{L,R}_{\theta}(x)$ are solutions of the corresponding Hamiltonians $ H_{L,R}(x)$, and take the following generic form in the two regions:

\begin{equation}
\begin{array}{ll}
\Psi^L_\theta(x)=
\frac{1}{\sqrt{2}} \begin{pmatrix}
c_{L} e^{i\theta}\\
i\\
-s_{L} e^{-i\theta}\end{pmatrix} e^{iq_x x} - \frac{r}{\sqrt{2}} \begin{pmatrix}
c_{L} e^{-i\theta}\\
-i\\
-s_{L} e^{i\theta}\end{pmatrix} e^{-iq_x x}, & x<0, \\
\Psi^{R}_\theta(x)= \frac{t}{\sqrt{2}} \begin{pmatrix}
c_{R} e^{i\theta}\\
i\\
-s_{R} e^{-i\theta}\end{pmatrix} e^{iq_x x} , & x>0.
\end{array}
\label{wave3_junction}
\end{equation}

To ensure the conservation of the probability current perpendicular to the interface, the wavefunction must satisfy an effective matching condition at the domain wall
\begin{equation}
 S_y^{R}\Psi_\theta^{R}(0^+)={\cal M}_{R L}S_y^{L}\Psi_\theta^{L}(0^-),
 \label{matching}
\end{equation}
with ${\cal M}_{R L}$ a $3\times3$ matching matrix that also satisfies
\begin{equation}
{\cal M}_{R L}S_y^{L}{\cal M}_{R L}^{\dagger}=S_y^{R}.
\label{matchsx}
\end{equation}
The solution to equation (\ref{matchsx}) is far from unique.
In the spirit of Ref. \cite{Romeo2018}, a natural physical Ansatz consists of choosing ${\cal M}_{R L}$
in the special linear group $SL(3,C)$ of $3\times3$ complex matrices of unit determinant.
With this constraint, it may be checked that a quite general solution
for the matching matrix ${\cal M}_{R L}$ takes the form
\begin{equation}
\mathcal{M}_{R L}= \begin{pmatrix}
 c_{RL} &  \lambda_2 c_{R} & s_{RL}\\
- \lambda_3  c_{L}& \lambda_1 &  \lambda_3 s_{L}\\
- s_{RL} & - \lambda_2 s_{R}& c_{RL}
\end{pmatrix},
\label{genematch matr}
\end{equation}
with
\begin{equation}
\begin{array}{l}
c_{R L}=c_{R}c_{L}+s_{L}s_{R},\\
s_{R L}=c_{L}s_{R}-c_{R}s_{L},
 \end{array}
 \label{crl_srl}
\end{equation}
such that $c_{RL}^2+s_{RL}^2=1$. Here, the parameters $\lambda_{i}$ are real-valued and must satisfy the constraint $\lambda_2 \lambda_3+\lambda_1=1$ which follows from the matching condition [Eq. (\ref{matchsx})].
Physically, we also expect that the matching matrix corresponds to the identity matrix when $\alpha_R=\alpha_L$; which implies that in this limit $\lambda_{2,3}=0$ and $\lambda_1=1$.
More generally, this constraint suggests that the phenomenological parameters $\lambda_i$ depend on $\alpha_R$, $\alpha_L$ and possibly on the energy $\varepsilon$.

\subsubsection{Domain wall with no interface {\it pseudo-potential}}

We first consider the Hamiltonian [Eq. (\ref{simple})] with no additional interface pseudopotential.
This case appears to describe well the tight-binding results
for the straight domain wall, as reported in Sec. \ref{sectionIII}.

As explained in more detail in the Appendix, in the absence of any interface potential, the  matching matrix is given by the product
$\mathcal{M}_{R L}= O_R O_L^{\dag}$ where $O_{\alpha}$, given in Eq. (\ref{o_alpha}), is the unitary transformation that {\it rotates} the spin basis according to $O^{\dagger}_\alpha S_y^\alpha O_\alpha=S_z$.
More explicitly, the matching matrix writes
\begin{equation}
\mathcal{M}_{R L}= \begin{pmatrix}
 c_{RL} & 0  &  s_{RL}\\
0 & 1 & 0\\
-s_{RL}& 0  & c_{RL}
\end{pmatrix}.
\label{match matr}
\end{equation}
This form of the matching matrix corresponds to $\lambda_{2,3}=0$ and $\lambda_{1}=1$.
In this case,
the eigenvalues of this matching matrix have unit modulus and are given by $(1,e^{\pm i \theta_{RL}})$
with $\tan  \theta_{RL}=\frac{s_{RL}}{c_{RL}}$.

More physically, applying the boundary condition [Eq. (\ref{matching})] with the latter simple matching matrix [Eq. (\ref{match matr})] and defining
\begin{equation}
 \Delta_{R L}=s^2_{R}-s^2_{L},
 \label{berryjump}
\end{equation}
which is proportional to the Berry phase jump $|\Phi_{L}-\Phi_{R}|/\pi$ (in unit of $\pi$) at the interface, we obtain the scattering amplitudes
\begin{subequations}
\begin{equation}
\begin{array}{l}
r=\dfrac{i\tan \theta \Delta_{R L}}{1- i\tan \theta \Delta_{R L}},
\end{array}
\label{amplir_junction}
\end{equation}
and
\begin{equation}
t=1+r=\dfrac{1}{1 - i \tan \theta \Delta_{R L}}.
\label{amplit_junction}
\end{equation}
\end{subequations}
Writing $r=|r|e^{-i\varphi_r}$ and $t=|t|e^{-i\varphi_t}$
we obtain the scattering phases $\varphi_r=\varphi_t-\pi/2$ with
\begin{equation}
\tan \varphi_t=  \Delta_{R L} \tan \theta.
\end{equation}
The transmission probability, within the continuum description, through the straight domain wall then reads
\begin{equation}
T_{\textrm{cont}}(\theta)=|t|^2=\dfrac{1}{1+ \tan^2 \theta \Delta_{R L}^2}.
\label{transstep}
\end{equation}

\begin{figure}[]
        \centering
      \includegraphics[width=0.5\textwidth]{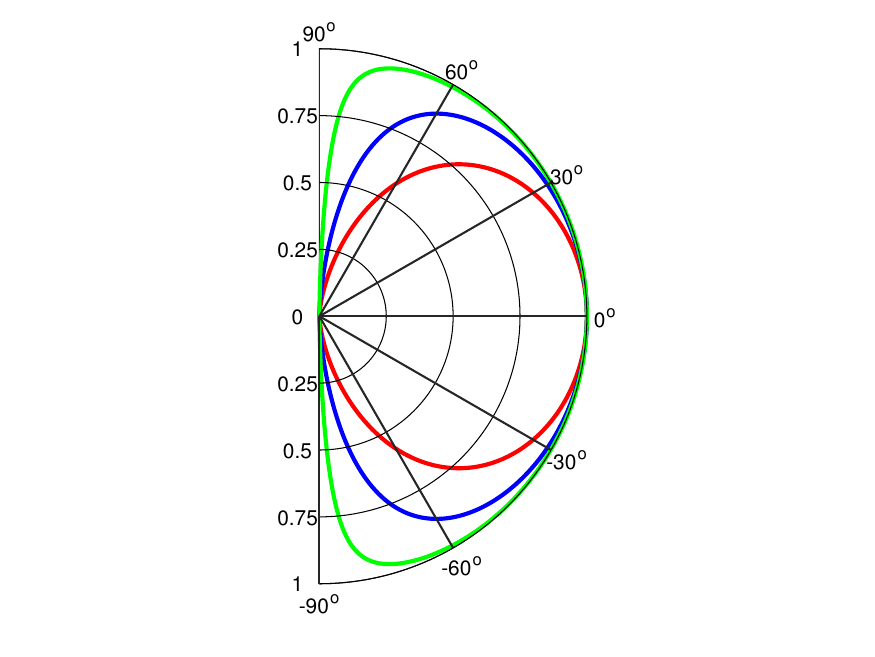} 
\caption{Transmission probability $T_{\textrm{cont}}(\theta)$ through the Berry phase domain wall
for increasing value of the Berry phase jump $\Delta_{R L}$:
green line $\Delta_{R L}=0.055$, blue line $\Delta_{R L}=0.22$ and red line $\Delta_{R L}=0.5$.}
\label{figure1_type1c}
\end{figure}

More physically, in contrast with conventional scattering by a potential barrier, the transmission probability $T_{\textrm{cont}}(\theta)$ does not depend on the energy $E$ of the incident Dirac particle, but only on the incident angle $\theta$. Similar to the case of an electrostatic barrier \cite{Allain2011,Illes2017}, the expression $T_{\textrm{cont}}(\theta)$ verifies $T_{\textrm{cont}}(\theta=0)=1$, indicating that a Klein tunneling effect occurs at normal incidence $\theta=0$ regardless of the  Berry phase jump at the domain wall interface. More unexpectedly,  we observe that $T_{\textrm{cont}}(\theta)$ exhibits the same angular dependence as the transmission probability obtained in the $\alpha-T_3$ model with a constant $\alpha$ and an electrostatic potential step $V(x)=V_0 \theta(x)$, where the step height $V_0=2E$ is equal to twice the energy of the incident Dirac particle. More quantitatively, in this latter case, the transmission probability is given by 
$T_{V_0=2E}(\theta)=[1+\tan^2 \theta (1-2s^2 _{\alpha})^2]^{-1}$ \cite{Illes2017}.

Figure \ref{figure1_type1c} shows the transmission probability $T_{\textrm{cont}}(\theta)$ as a function of the incidence angle $\theta$ for several values of the Berry phase jump $\Delta_{R L}$. 
Figure \ref{equiv type2} provides a schematic comparison between scattering through the Berry phase domain wall (left) and scattering through a potential step in the $\alpha-T_3$ model with step height $V_o=2E$ (right).

The above results for a single Berry phase domain wall suggest that, in the presence of two Berry phase domain walls separated by a finite distance, Fabry–Pérot–like resonance phenomena may arise. For multiple domain walls with random Berry phases jumps, an interesting perspective would be to explore possible localization effects.

To finish with, for electronic systems, a more directly measurable quantity is the ballistic conductance, which is given by using the Landauer-B\"uttiker formula \cite{Blanter2000} 
\begin{equation}
G =\frac{4e^2}{h}\sum_{k_y}T_{\textrm{cont}}(k_y)
\label{conductance1}
\end{equation}
where the factor of $4$ accounts for spin and valley degeneracy. This expression can be evaluated explicitly as
\begin{equation}
G_{RL}=G_{\alpha\alpha} \left[ a_{RL}^2+(a_{RL}-a_{RL}^3)\operatorname{artanh}(1/a_{RL})\right]
\label{cond_cont}
\end{equation}
where $G_{\alpha\alpha}=\frac{4e^2}{\pi h}\frac{L_y E}{\hbar v_F}$ is the conductance in the case of the perfect transmission (i.e., when $\alpha_R=\alpha_L$), $a_{RL}=(1-\Delta_{RL}^2)^{-1/2}$ and $L_y$ is the width of the sample in the $y$ direction.

\subsubsection{Domain wall with an interface pseudo-potential}

As explained in the Appendix, in the rotated bases, by adding an interface pseudopotential term ${\bar V}(x)=\delta(x) {\bar V}$, with ${\bar V}= i \hbar v_F A S_z$ and $A$ a $3\times3$ matrix,
then to leading order,  the effective matching matrix writes ${\cal M}_{R L}=O_R (1+A)O_L^{\dag}$.
It is then immediate to check that by taking
\begin{equation}
A=
\begin{pmatrix}
a&0&b\\
0&0&0\\
-b^*&0&-a^*
\end{pmatrix},
\end{equation}
with
\begin{equation}
 \begin{array}{l}
  a=\frac{1}{2}(-1+\lambda_1+ i(\lambda_2+\lambda_3))\\
  b=\frac{1}{2}(1-\lambda_1+ i(\lambda_2-\lambda_3)),
 \end{array}
\end{equation}
one recovers the matching matrix generic form given in Eq. (\ref{genematch matr}).

Anticipating on Sec. \ref{sectionIII}, we further consider the particular case, $\lambda_2=0$ which imply $\lambda_1=1$.
In this situation, the matching matrix takes the form
\begin{equation}
\mathcal{M}_{R L}= \begin{pmatrix}
 c_{RL} & 0 &  s_{RL}\\
- \lambda_3  c_{L}& 1 &  \lambda_3 s_{L}\\
- s_{RL} &0 & c_{RL}
\end{pmatrix}.
\label{genematch matr3}
\end{equation}
Applying the matching condition [Eq. (\ref{matching})], we then obtain the scattering amplitudes
\begin{equation}
 \begin{array}{ll}
r&=\dfrac{i(\Delta_{RL}\sin \theta+ \lambda_3/2)}
{\cos \theta-i(\Delta_{RL}\sin \theta+ \lambda_3/2)},\\
t&=\dfrac{\cos \theta}{\cos \theta-i(\Delta_{RL}\sin \theta+\lambda_3/2)}.\\
 \end{array}
\end{equation}
The transmission probability, within the continuum description, through the zigzag domain wall is given by
\begin{equation}
T_{\textrm{cont}}(\theta)=\frac{\cos^2 \theta}{\cos^2 \theta+(\Delta_{RL}\sin \theta+\lambda_3/2)^2}. 
 \label{cont type4}
\end{equation}
This last expression will appear useful to describe the tight-binding results
of the zigzag domain wall, as reported in the Sec. \ref{sectionIII}.

\section{Lattice description}
\label{sectionIII}

In this section, we revisit  the problem of scattering through a Berry phase domain wall using the tight-binding description of the $\alpha-T_3$ model on the dice lattice. Within this framework, we investigate two kinds of domain walls: a straight (see Fig. \ref{figtype1}) and a zigzag (see Fig. \ref{figtype2}) domain walls.

\subsection{Homogeneous tight-binding  $\alpha-T_3$ model}

We start by reminding the $\alpha-T_3$ nearest neighbors tight-binding model on the dice lattice \cite{Raoux2014}. The dice lattice is a triangular Bravais lattice with three sublattices $X=A,B,C$.
We denote  ${\bm a}_1=a(\frac{3}{2},\frac{\sqrt{3}}{2})$, ${\bm a}_2=a(0,\sqrt{3})$ the Bravais lattice vectors with $a=1$ the intersite distance.
To shorten the writings, we adopt a {\em second-quantized} notation and define $X^{\dagger}_{n,m}\equiv |X_{n,m}\rangle$ the atomic basis state at position ${\bm r}_X+n  \bm{a}_1+m  \bm{a}_2$
with ${\bm r}_A=0,{\bm r}_B=\frac{2}{3}{\bm a}_1-\frac{1}{3}{\bm a}_2$
and ${\bm r}_C=\frac{1}{3}({\bm a}_1+{\bm a}_2)$.

The homogeneous $\alpha-T_3$ nearest-neighbor tight-binding Hamiltonian then writes as
\begin{equation}
 \begin{array}{ll}
H_\alpha&=  c_\alpha t\sum_{m,n} A^{\dagger}_{n,m}(B_{n,m}+B_{n-1,m}+B_{n-1,m+1})+\textrm{H.c}\\
& +s_\alpha t\sum_{m,n}  C^{\dagger}_{n,m}(B_{n,m}+B_{n,m+1}+B_{n-1,m+1}) +\textrm{H.c},
\end{array}
\end{equation}
such that each $B$ site is coupled to three $A$ sites via the hopping amplitude $c_\alpha t$ and to three $C$ sites via the hopping amplitude $s_\alpha t$;
with $c_\alpha=(1+\alpha^2)^{-1/2}$, $s_\alpha=\alpha(1+\alpha^2)^{-1/2}$, and $t$ is a characteristic energy scale.

We now remind the specific $3\times3$ Bloch Hamiltonian matrix within our notation.
More concretely, defining the sublattices Bloch basis of creation operators
$X^{\dagger}({\bm k})=e^{i {\bm k}\cdot {\bm r_X}}\sum_{n,m}
e^{i {\bm k}\cdot(n\bm{a}_1+m \bm{a}_2)}X^{\dagger}_{n,m}$,
the corresponding Bloch Hamiltonian matrix takes the form
\begin{equation}
H_\alpha({\bm k})=t\left(
 \begin{array}{ccc}
  0& c_\alpha f({\bm k})&0\\
  c_\alpha f^*({\bm k})&0&s_\alpha f({\bm k})\\
  0&s_\alpha f^*({\bm k})&0
 \end{array}\right),
  \label{hblock}
 \end{equation}
 with $f({\bm k})=e^{i k_x}+2e^{-i \frac{k_x}{2}}\cos{\frac{\sqrt{3}k_y}{2}}$.
With this form of $f({\bm k})$, the two Dirac points are at ${\bm K}_\pm=(0,\pm \frac{4\pi}{3\sqrt{3}})$.
Writing ${\bm k}={\bm K}_\pm+ {\bm q}$ and expanding to linear order in ${\bm q}$ around the two Dirac valleys, one obtains the effective low-energy model Hamiltonian given by Eq. (\ref{Hamilton0}), with $\hbar v_F=\frac{3t}{2}$.

We also remind that the Bloch state eigenfunctions take the form
\begin{equation}
\psi_{\pm,\theta}(\vec{r})=\frac{1}{\sqrt{2}}\begin{pmatrix}
c_{\alpha} \frac{ f({\bm k})}{|f({\bm k})|}\\
\pm 1\\
s_{\alpha} \frac{ f^*({\bm k})}{|f({\bm k})|}
\end{pmatrix} e^{i\vec{q}\cdot \vec{r}},
\hspace{0.1cm}
\psi_{0,\theta}(\vec{r})=\begin{pmatrix}
-s_{\alpha} \frac{ f({\bm k})}{|f({\bm k})|}\\
0\\
c_{\alpha} \frac{ f^*({\bm k})}{|f({\bm k})|}
\end{pmatrix} e^{i\vec{q}\cdot \vec{r}}.
\label{psi1}
\end{equation}
Similarly to their low-energy counterparts [Eq. (\ref{psi0})], the expressions in Eq. (\ref{psi1}) suggest that the components on sublattices $A$ and $C$ encode all the dependencies on the parameter $\alpha$ and on the wavevector ${\bm k}$. In contrast, the $B$ component is constant. We believe that this property is at the origin of the strong difference between the scattering probability through the straight domain wall and that through the zigzag domain wall, as reported in the following sections.

\subsection{Tight-binding models of a Berry phase domain wall}

\begin{figure}[]
\setlength{\unitlength}{1mm}
\includegraphics[width=0.5\textwidth]{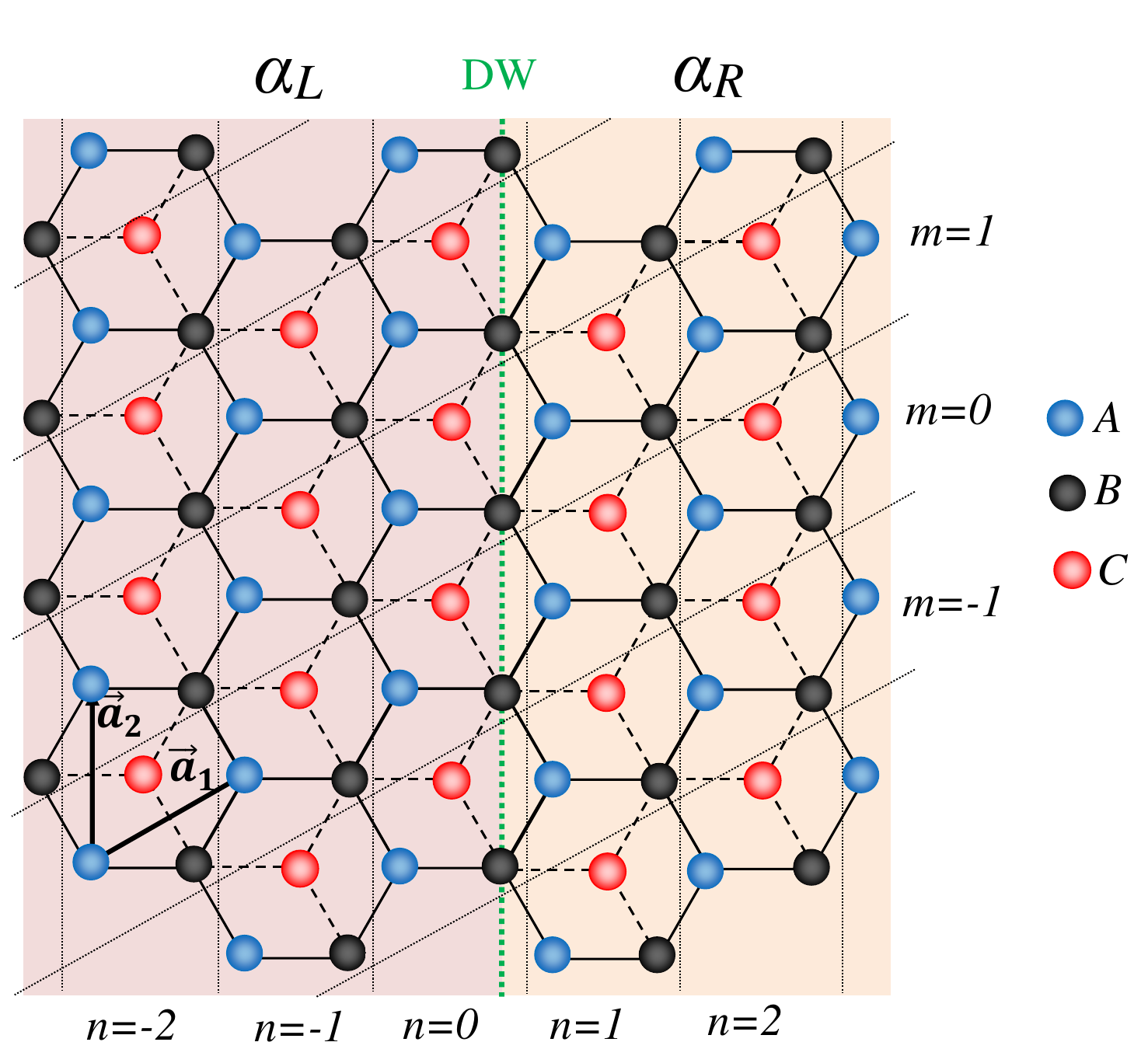}
\caption{Lattice description of the {\it straight} domain wall.
Bravais lattice vectors ${\bm a}_1,{\bm a}_2$. The green dotted line is the virtual domain-wall line (parallel to ${\bm a}_2$) that separates the regions $\alpha=\alpha_L$ for $n\le 0$ and $\alpha=\alpha_R$ for $n>0$. In each region, the nearest-neighbor hopping amplitudes are $t_{AB}=c_{\alpha}t$ (black continuous line) and $t_{CB}=s_{\alpha}t$ (black dashed line).}
\label{figtype1}
\end{figure}

We now consider inhomogeneous Hamiltonian models describing an interface
parallel to the ${\bm a}_2$ axis ($y$-axis), located at position $n=0$.
This interface separates two domains characterized by different values of the parameter $\alpha$:
 $\alpha_L$ on the left side ($n < 0$) and $\alpha_R$ on the right side ($n > 0$).
With this choice of domain-wall orientation on the lattice, there are two possible kinds of domain wall.
The first kind of domain wall is a virtual {\it straight line} composed only of $B$ sites at positions ${\bm r}_B+m \bm{a}_2$ (and $n=0$), as illustrated in figure  (see Fig. \ref{figtype1}).
In this straight domain-wall configuration, each of the $B$ sites on the domain wall sees an inhomogeneous environment that mixes $\alpha_R$ and $\alpha_L$. In contrast, the $A$ and $C$ sublattices retain a homogeneous local environment, with each site coupled exclusively to either $\alpha_L$ or $\alpha_R$.
More concretely, for this {\it straight} domain wall, the inhomogeneous Hamiltonian
simply writes $H=H_{L}(n< 0)+H_{RL}(n=0)+H_{R}(n> 0)$,
with the interface contribution $H_{R L}$ that writes
\begin{equation}
 \begin{array}{ll}
H_{R L}&= t \sum_{m}  c_L A^{\dagger}_{0,m}( B_{-1,m}+ B_{-1,m+1}+ B_{0,m})+\textrm{H.c}\\
& +t \sum_{m} s_L C^{\dagger}_{0,m}( B_{-1,m+1}+ B_{0,m}+ B_{0,m+1})+  \textrm{H.c}.
\end{array}
\label{interfacepotential2}
\end{equation}
As we explain in detail in the following sections, the scattering properties of this straight domain wall are in good quantitative agreement with those predicted by the simplest continuum model, with no additional interface  pseudopotential.

\begin{figure}[h!]
\setlength{\unitlength}{1mm}
\includegraphics[width=0.5\textwidth]{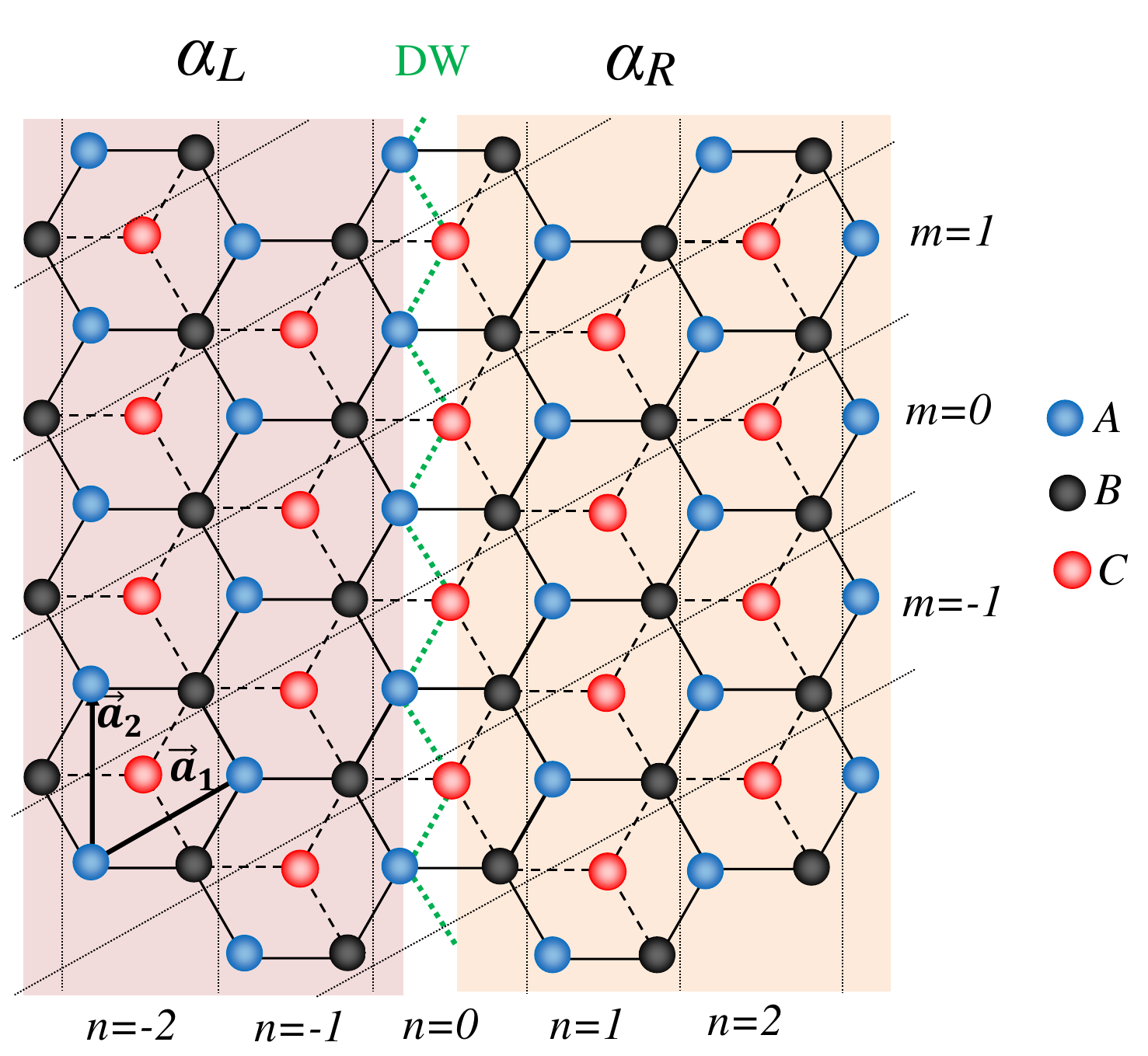}
\caption{Lattice description of the {\it zigzag} domain wall. The green dotted line is the virtual domain-wall line.}
\label{figtype2}
\end{figure}
The second kind of domain wall is a virtual {\it zigzag line}, composed of $A$ and $C$ sites at positions ${\bm r}_A+m \bm{a}_2$ and  ${\bm r}_C+m \bm{a}_2$ (and $n=0$), as depicted in Fig. \ref{figtype2}.
In contrast with the straight domain wall, each $A$ and $C$ site on the zigzag domain wall sees an inhomogeneous local environment that mixes $\alpha_R$ and $\alpha_L$, while each $B$ site sees either $\alpha_L$ or $\alpha_R$.
For this second kind of domain wall, the inhomogeneous Hamiltonian  takes the form
$H=H_{L}(n< 0)+H_{R L}(n=0)+H_{R}(n>0)$
where the {\em interface} contribution $H_{R L}$ is given by
\begin{equation}
 \begin{array}{ll}
H_{R L}&= t \displaystyle \sum_m  A^{\dagger}_{0,m}(c_L B_{-1,m}+c_L B_{-1,m+1}+c_R B_{0,m})+\textrm{H.c}\\
& +t \displaystyle \sum_m C^{\dagger}_{0,m}(s_L B_{-1,m+1}+s_R B_{0,m}+s_R B_{0,m+1})+  \textrm{H.c}.
\end{array}
\label{interfacepotential2}
\end{equation}
This distinct interface contribution has a significant impact on the transmission probability.  
As we show below, it is nevertheless possible to recover the tight-binding transmission probability in the low-energy limit by using a continuum approach with an effective matching matrix, as in Eq. (\ref{genematch matr}), with some specific values for the parameter $\lambda_i$. In other words, the low-energy properties of this zigzag domain wall may be qualitatively described by a continuum model with a specific interface pseudopotential.

We believe that the pronounced difference in scattering properties between the two types of domain walls can be explained as follows. 
The straight domain wall consists only of $B$ sites and therefore mostly affects the wavefunction amplitudes on these sites. 
 However, as seen from both the Bloch and scattering states [see Eqs.~(\ref{psi0}), (\ref{wave3_junction}), (\ref{psi1})], the $B$ component is constant and does not depend on either the parameter $\alpha$ or the wavevector ${\bm k}$. As a result, it carries little to no essential information about the state. In contrast, the zigzag domain wall involves $A$ and $C$ sites, and thus scattering off this domain wall directly impacts the $A$ and $C$ components, which encode all the dependence of the Bloch (scattering) states on $\alpha$ and ${\bm k}$.

\begin{figure}[h!]
\setlength{\unitlength}{1mm}
\includegraphics[width=0.5\textwidth]{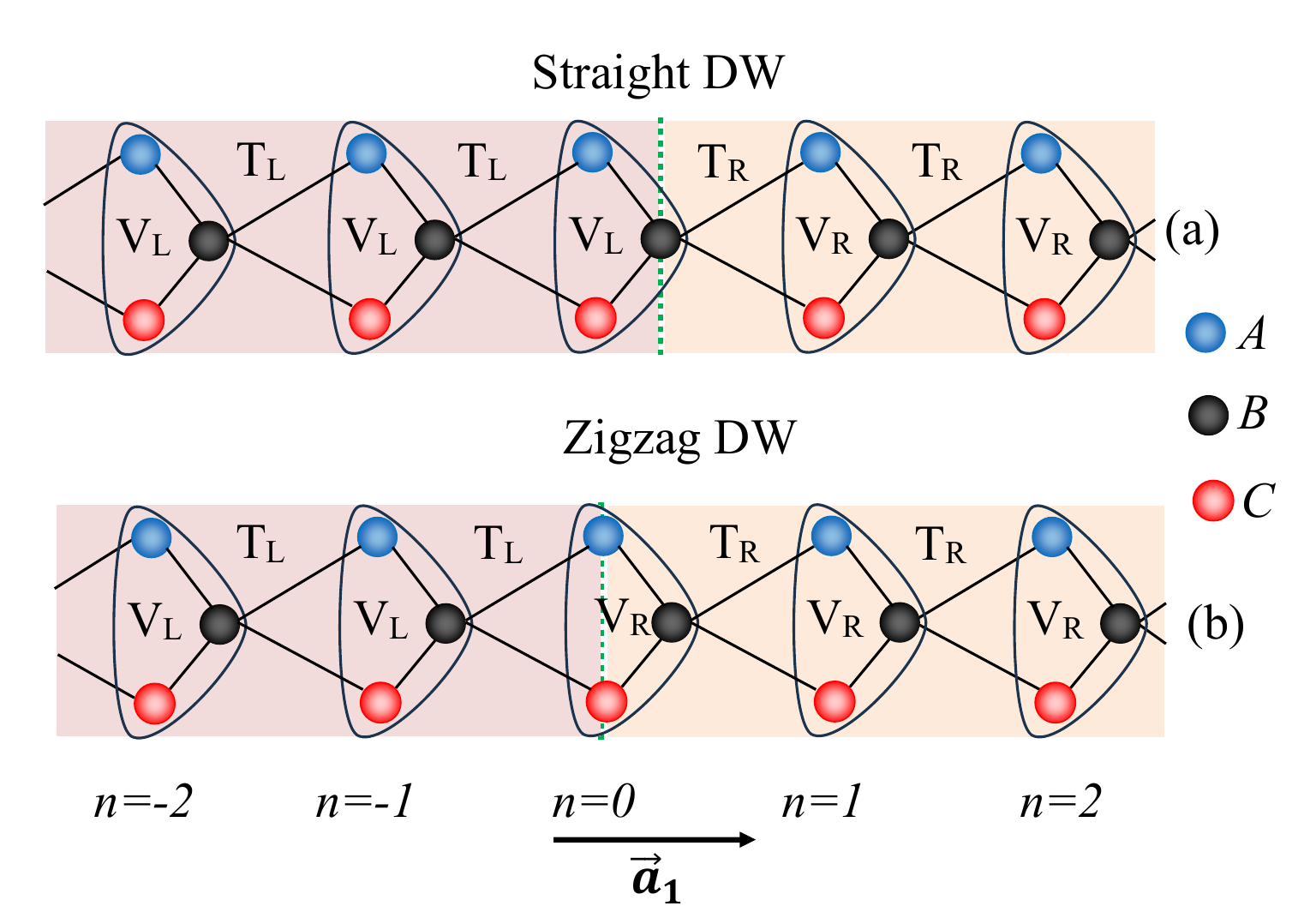}
\caption{Effective one-dimensional model picture of the straight (a) and zigzag (b) domain walls.}
\label{fig1d}
\end{figure}

The previous models are translation invariant along the direction $\bm{a}_2=\sqrt{3} {\bm e}_y$ of the domain wall. We can thus use Bloch's theorem along $\bm{a}_2$ and define, for each sublattice,  a one-dimensional Bloch basis of creation operators
$X^{\dagger}_{n}(k_y)=
 \sum_m e^{i m{\bm k\cdot \bm a_2}}X^{\dagger}_{n,m}$, where $X = A, B, C$.
Defining further the three component creation operators $L_{n}^{\dagger}(k_y)\equiv (A^{\dagger}_{n},B^{\dagger}_{n},C^{\dagger}_{n})$,
the  homogeneous  $\alpha-T_3$ model rewrites as $H_\alpha=\int_{BZ} \dfrac{dk_y}{2\pi} H_\alpha(k_y)$, where $H_\alpha(k_y)$ defines a $k_y$-dependent effective one-dimensional tight-binding model, which takes the form
\begin{equation}
H_\alpha(k_y)=\sum_n L_{n}^{\dagger}V_\alpha L_{n}
+L_{n}^{\dagger }T_\alpha L_{n-1} +L_{n-1}^{\dagger}T_\alpha ^\dagger L_{n},
\end{equation}
 with a pseudopotential matrix
\begin{equation}
V_\alpha(k_y)=t\left(
 \begin{array}{ccc}
  0& c_\alpha &0\\
  c_\alpha&0&s_\alpha(1+z_2^*)\\
  0&s_\alpha(1+z_2)&0
 \end{array}\right),
  \label{valpha}
 \end{equation}
and an effective nearest-neighbor hopping matrix
\begin{equation}
T_\alpha(k_y)=t\left(
 \begin{array}{ccc}
  0& c_\alpha (1+z_2)&0\\
  0&0&0\\
  0&s_\alpha z_2 &0
 \end{array}\right),
\end{equation}
where $z_2=e^{i{\bm k\cdot \bm a_2}}=e^{i {\sqrt{3}k_y}}$.
For the straight domain wall, the effective one-dimensional Hamiltonian takes the form
$ H(k_y)=H_{L}(k_y, n\le 0)+H_{R}(k_y, n> 0)$,
as schematically illustrated in Fig \ref{fig1d} (a). 
In this case, the pseudopotential at the interface site $n=0$ is given by $V_L$. 
In contrast, for the zigzag domain wall, the corresponding effective Hamiltonian can be represented as in Fig \ref{fig1d} (b), but with the pseudopotential at $n=0$  now given by $V_R$.
Within this effective one-dimensional model, the key difference between the straight and zigzag domain walls lies in the pseudopotential acting at the interface site $n=0$.
 We believe that this explains why we find that the two domain walls differ by an effective interface pseudopotential effect in their continuum description.
 Furthermore, we note that this interface potential should be related to $V_R(k_y)-V_L(k_y)$, which depends on the difference $\alpha_R-\alpha_L$ and also on the transverse momentum $k_y$.
We can therefore expect that, in the continuum limit, the effective interface pseudopotential depends on both $\alpha_R-\alpha_L$ and the transverse momentum $q_y$.

An important property of each of these effective inhomogeneous Hamiltonians model is that they anticommute (for each $k_y$) with the chiral symmetry operator $S=\sum_n L_{n}^{\dagger} S L_{n}$ with $S=\textrm{diag}(1,-1,1)$. As a consequence, the flat band at zero energy remains whatever the spatial inhomogeneity in the $\alpha$ parameter.

\subsection{Scattering properties of the straight domain wall}
\label{scatt-type1}

We now determine the scattering properties of the lattice model for the straight domain wall (see Fig. \ref{figtype1}).
More quantitatively, we look for scattering states $\Psi^{\dagger}(\varepsilon,k_y)$ of energy $E=\varepsilon t$ and momentum $k_y$ that write
$\Psi^{\dagger}=\sum_n \sum_{X}\psi_{n}^X(\varepsilon,k_y) X^{\dagger}_{n}(k_y)$
($X=A,B,C$) and
that verify the eigenvector constraint $H(k_y)\Psi^{\dagger}=t\varepsilon \Psi^{\dagger}$.
More explicitly, the latter equality-constraint translates into the following systems of equations for the scattering state amplitudes $\psi_n^X(\varepsilon,k_y)$
\begin{equation}
 \begin{array}{ll}
\begin{array}{l}
 n<0, \alpha=L\\
n>0, \alpha=R\\
\end{array}
  &\left\{
 \begin{array}{l}
  \varepsilon \psi_{n}^A=c_\alpha [\psi_{n}^B +(1+z_2)\psi_{n-1}^B],\\
  \varepsilon \psi_{n}^C=s_\alpha [(1+z_2)\psi_{n}^B+z_2\psi_{n-1}^B],\\
  \begin{split}
  \varepsilon \psi_{n}^B=&c_\alpha [\psi_{n}^A +(1+z_2^*)\psi_{n+1}^A]+\\&s_\alpha  [(1+z_2^*)\psi_{n}^C +z_2^*\psi_{n+1}^C],
\end{split}  
  \\
 \end{array}\right.\\
& \\
n=0 &\left\{
 \begin{array}{l}
  \varepsilon \psi_{0}^A=c_{L}[ \psi_{0}^B+(1+ z_2) \psi_{-1}^B],\\
  \varepsilon \psi_{0}^C=s_{L}[(1+z_2)\psi_{0}^B+z_2\psi_{-1}^B],\\
  \begin{split}
  \varepsilon \psi_{0}^B=&c_{L} \psi_{0}^A + c_{R}(1+z_2^*)\psi_{1}^A+\\&s_{L} (1+z_2^*)\psi_{0}^C+s_{R}z_2^*\psi_{1}^C,
  \end{split}  
  \\
 \end{array}\right.\\
 \end{array}
 \label{eq 2}
\end{equation}
From this point, it appears more convenient to return to first quantized notation and define $|\Psi_{n}\rangle \equiv \left(\psi_{n}^A, \psi_{n}^B, \psi_{n}^C\right)^T$. Interestingly (thanks to chiral symmetry), we can rewrite the full set of equations for $n<0$, $n=0$, and $n>0$  in terms of three  distinct transfer matrices as
\begin{equation}
\begin{array}{ll}
 n \le 0 & | \Psi_{n}\rangle= M_{L}| \Psi_{n-1}\rangle,\\
   & |\Psi_{1}\rangle= M_{RL}| \Psi_{0}\rangle,\\
 n>1 &  |\Psi_{n}\rangle= M_{R} |\Psi_{n-1}\rangle.
 \end{array}
 \label{transfer2}
\end{equation}
This system of equations implies that somehow the slice $n=0$ can be considered on the $L$ region.
More quantitatively, the transfer matrices $M_{\alpha}$ ($\alpha=R$ or $\alpha=L$) and $M_{R L}$  are non-Hermitian and write
\begin{equation}
\begin{split}
M_\alpha(\varepsilon,k_y)=
&\left(
 \begin{array}{ccc}
        \varepsilon &-c_\alpha & 0\\
        -c_\alpha(1+z_2^*) &0&- s_\alpha z_2^* \\
        0&-s_\alpha (1+z_2)  &\varepsilon
\end{array}
\right)^{-1}\\& \times
 \left(
     \begin{array}{ccc}
        0&c_\alpha (1+z_2) & 0 \\      
        c_\alpha & -\varepsilon&s_\alpha(1+z_2^*)\\
        0& s_\alpha z_2  & 0
       \end{array}
       \right),
       \end{split}
       \label{trans mat}
\end{equation} 
and
\begin{equation}
\begin{split}
M_{RL}(\varepsilon,k_y)=
&\left(
 \begin{array}{ccc}
        \varepsilon &-c_{R} & 0\\
        -c_{R}(1+z_2^*) &0&- s_{R} z_2^*  \\
          0&-s_{R} (1+z_2)  &\varepsilon\\
\end{array}
\right)^{-1}\\& \times
 \left(
     \begin{array}{ccc}
        0 &c_{R} (1+z_2)& 0 \\
        c_{L} &-\varepsilon& s_{L}(1+z_2^*)\\
        0& s_{R} z_2  & 0\\
       \end{array}
       \right),
       \end{split}
       \label{trans mat typ2}
\end{equation} 
Remarkably, the matrices $M_{\alpha}$ with $\alpha=L,R$
have eigenvalues $\lambda_s(\varepsilon,k_y)$ ($s=\pm,0$) that are independent of $\alpha$
\begin{equation}
\lambda_\pm(\varepsilon,k_y)= \frac{1}{2(1+z_2)}[\gamma \pm \sqrt{\gamma^2-4z_2(1+z_2)^2}], \ \ \  \lambda_{0}=0,
\end{equation}
with $\gamma(\varepsilon,k_y)=z_2 \varepsilon^2-(z_2^2+3z_2+1)$ such that $\lambda_+ \lambda_-=z_2$.
For each $\alpha$, the transfer matrices $M_{\alpha}$, being non-Hermitian, so one needs to define so-called {\em right} eigenvectors $|\Psi^{\alpha}_s\rangle$ and {\em left} eigenvectors
$|\Psi^{\alpha}_s\rrangle$, which satisfy $M_{\alpha}|\Psi^{\alpha}_s\rangle=\lambda_s |\Psi^{\alpha}_s\rangle$ and $M_{\alpha}^{T}|\Psi^{\alpha}_s\rrangle^*=\lambda_s |\Psi^{\alpha}_s\rrangle^*$, with the condition $\llangle \Psi^{\alpha}_s|\Psi^{\alpha}_{s'}\rangle={\cal N}_s\delta_{ss'}$.
Accordingly, the matrix $M_{\alpha}$ can be decomposed as $M_{\alpha}=\sum_s \frac{\lambda_s}{{\cal N}_s} |\Psi^{\alpha}_s\rangle \llangle \Psi^{\alpha}_s|$, where the normalization constants are ${\cal N}_0=z_2(1+z_2)$ and ${\cal N}_\pm= (z_2-\lambda_\pm ^2)(1+z_2)$.
More quantitatively, one obtains
\begin{equation}
\begin{array}{cc}
  |\Psi^{\alpha}_0\rangle=
  \left(
  \begin{array}{c}
  -(1+z_2) s_\alpha\\
  0\\
  z_2 c_\alpha
 \end{array}
 \right),\\
   |\Psi^{\alpha}_\pm\rangle=
  \left(
  \begin{array}{c}
   c_\alpha(\lambda_\pm+1+z_2)\\
  \varepsilon \lambda_\pm\\
   s_\alpha((1+z_2) \lambda_\pm+z_2)\\
 \end{array}
 \right).
 \end{array}
\end{equation}
and
\begin{equation}
\begin{split}
\begin{array}{cc}
  &|\Psi^{\alpha}_0\rrangle^*=
  \left(
  \begin{array}{c}
  -z_2 s_\alpha\\
  -s_\alpha c_\alpha (1+z_2 +z_2^2)/\varepsilon\\
 (1+z_2) c_\alpha 
 \end{array}
 \right),\\
  &|\Psi^{\alpha}_\pm\rrangle^*=
  \left(
  \begin{array}{c}
   c_\alpha z_2 \\
  -(c_\alpha^2 z_2 +s_\alpha^2 (1+z_2)^2+(1+z_2)\lambda_\pm)/\varepsilon\\
   s_\alpha(1+z_2) \\
 \end{array}
 \right).
 \end{array}
 \end{split}
\end{equation}
Note that, in contrast with the eigenvalues $\lambda_{s}$, the eigenvectors $|\Psi^{\alpha}_{s}\rangle$ explicitly depend on $\alpha$.
For later use, we denote the norms as $N_{\pm}^{\alpha}=\langle \Psi^{\alpha}_\pm|\Psi^{\alpha}_\pm \rangle=2\varepsilon^2$, which are independent of 
$\alpha$, whereas $N_{0}^{\alpha}=\langle \Psi^{\alpha}_0|\Psi^{\alpha}_0 \rangle=1+(1+z_2+z_2^*)s_\alpha^2$.

Choosing the energy $\varepsilon$ in the band spectrum by writing
\begin{equation}
\varepsilon=\left|1+z_1+z_2\right|,
\label{energy}
\end{equation}
with $z_1=e^{i {\bm k} \cdot {\bm a}_1}$, it is straightforward to verify that the scattering eigenvalues $\lambda_{\pm}$ are pure phases
\begin{equation}
 \lambda_\pm(\varepsilon,k_y)=e^{i\left( \frac{ \sqrt{3}}{ 2} k_y\pm \frac{3}{ 2}k_x\right )}.
\end{equation}
Accordingly, the corresponding eigenstates $|\Psi^{\alpha}_{\pm}\rangle$ can be interpreted as itinerant right- and left-moving plane-wave eigenvectors, whereas $|\Psi^{\alpha}_{0}\rangle$ is obviously a nonitinerant eigenvector, which can only exist on the slice $n=0$.

A generic scattering state, of energy $\varepsilon t$ and momentum $k_y$, may then be written as
$|\Psi\rangle=\sum_n |\Psi_n\rangle$,
with
\begin{equation}
 |\Psi_n\rangle=\left \lbrace
 \begin{array}{lr}
  \sum_{s=\pm} a_s \lambda_s ^n |\Psi_s^{L}\rangle & n< 0,\\
  \sum_{s=\pm,0} a_s \lambda_s ^n |\Psi_s^{L}\rangle & n=0,\\
  \sum_{s=\pm} b_s \lambda_s ^n |\Psi_s^{R}\rangle & n> 0.
 \end{array}\right.
\end{equation}
This form accounts for the fact that the eigenmode $|\psi_0^{\alpha}\rangle$ is nonitinerant and thus can only appear on the slice $n=0$.
 Since  the layer $n=0$ belongs to the $L$ region for the considered domain wall, only $a_0$ might be nonzero while $b_0=0$. 
The scattering amplitudes $a_s$  and  $b_s$ are constrained by the interface boundary condition $|\Psi_{1}\rangle= M_{RL}| \Psi_{0}\rangle$. 
More concretely, multiplying both sides of this equation by the left eigenvectors $\llangle \Psi^{R}_{s'}|$ yields
\begin{equation}
 {\cal N}_{s'} \lambda_{s'} b_{s'}=\sum_{s=\pm,0}{M}_{s's} a_s,
\end{equation}
with $s'=\pm$ and
\begin{equation}
 {M}_{s's}=\llangle \Psi^{R}_{s'}|M_{R L}|\Psi^{L}_s\rangle.
\end{equation}

More quantitatively, we obtain:
\begin{equation}
 \begin{array}{l}
M_{++}=[(1+z_2+z_2^2)\Delta_{R,L}-(1+z_2)(\lambda_+-\lambda_-)]\lambda_+^2,\\
M_{--}=[(1+z_2+z_2^2)\Delta_{R,L}+(1+z_2)(\lambda_+-\lambda_-)]\lambda_-^2,\\
M_{+-}=M_{-+}=z_2(1+z_2+z_2^2)\Delta_{R,L},\\
M_{+0}=M_{-0}=M_{00}=0.
 \end{array}
\end{equation}
with $\Delta_{RL}=s^2_{R}-s^2_{L}$ denoting the Berry phase jump (in unit of $\pi$) at the interface.
These relations imply that $a_0$ plays no role and can therefore be set to zero.

For an incident right-moving wave of amplitude $a_+$ in region $L$, the reflected left-moving wave has amplitude $a_-=r a_+$ and the transmitted right-moving wave in region $R$ has amplitude $b_+=t a_+$ (we impose $b_-=0$ since a finite $b_-$ would represent an incident left-moving wave in region $R$) such that we obtain
\begin{equation}
\begin{array}{lll}
 r=-\frac{{M}_{-+}}{{M}_{--}}& \textrm{and} & t=\frac{{M}_{++}{M}_{--}-{M}_{-+}{M}_{+-}}{\lambda_+{\cal N}_{+}{M}_{--}},
\end{array}
\end{equation}
or more explicitly
\begin{equation}
 \begin{array}{l}
r=\frac{-\lambda_+}{\lambda_-}\dfrac{i\Gamma\Delta_{RL}}{1+i\Gamma\Delta_{RL}},\\
\textrm{and}\\
t=\dfrac{1}{1+i\Gamma\Delta_{RL}},
 \end{array}
 \label{t type1}
\end{equation}
where
\begin{equation}
 \Gamma(\varepsilon,k_y)=-i\dfrac{(1+z_2+z_2^2)}{(1+z_2)(\lambda_+-\lambda_-)}=
 \frac{\frac{1}{4}-\cos^2\frac{\sqrt{3}}{2}k_y}{\cos\frac{\sqrt{3}}{2}k_y\sin\frac{3}{2}k_x}.
 \label{Gamma}
\end{equation}
The longitudinal momentum $k_x$ is obtained from the expression of the energy $\varepsilon$ [Eq. (\ref{energy})] as
\begin{equation}
k_x=\frac{2}{3}\arccos\frac{\varepsilon^2-4\cos^2\frac{\sqrt{3}}{2}k_y-1}{4\cos\frac{\sqrt{3}}{2}k_y}.
\label{kx_e_ky}
\end{equation}
Finally, for the straight domain wall, we deduce that the transmission probability within the lattice model is given by
\begin{equation}
T_{\textrm{latt}}(\varepsilon,k_y)=\frac{1}{1+\Gamma^2 \Delta_{RL}^2}.
\label{T type1}
\end{equation}

Interestingly, for this straight domain wall,
the expressions for $r, t$ and $T_{\textrm{latt}}$ in Eqs. (\ref{t type1}), (\ref{T type1}) of the lattice model are in direct correspondence with 
those given in Eqs. (\ref{amplir_junction}), (\ref{amplit_junction}), and (\ref{transstep}) obtained for the continuous model (up to a phase for $r$)
with $\Gamma(\varepsilon,k_y)$ in the lattice model substituted for $\tan \theta$ in the continuous model.
\begin{figure}[]
\setlength{\unitlength}{1mm}
\includegraphics[width=0.5\textwidth]{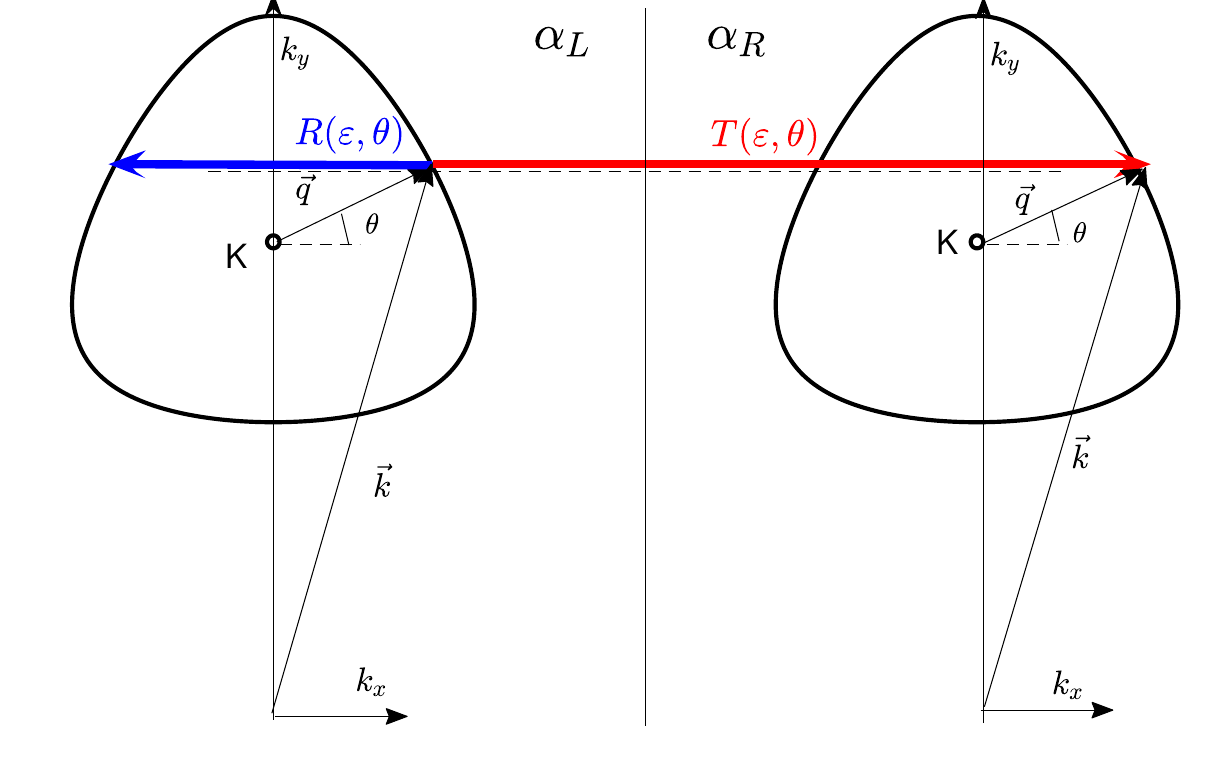}
\caption{Constant energy $\varepsilon<1$ contour around the Dirac point $K(0,\frac{4\pi}{3\sqrt{3}})$ and schematic representation of the scattering through the $\alpha_L$-$\alpha_R$ domain wall for a given energy $\varepsilon$ and incidence angle $\theta$. }
\label{fig_contour}
\end{figure}

\begin{figure}[]
        \centering
      \includegraphics[width=0.4\textwidth]{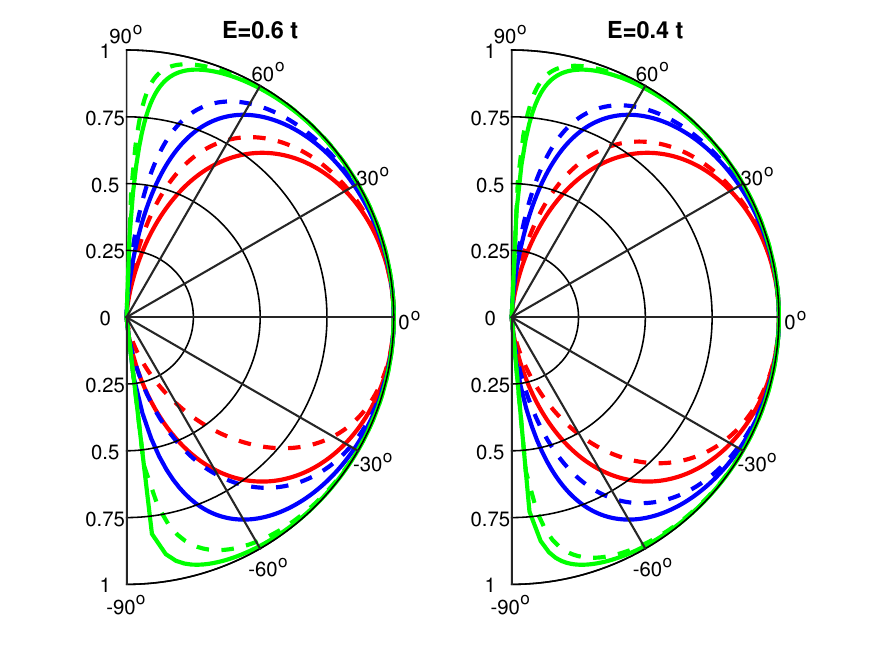}
      \includegraphics[width=0.4\textwidth]{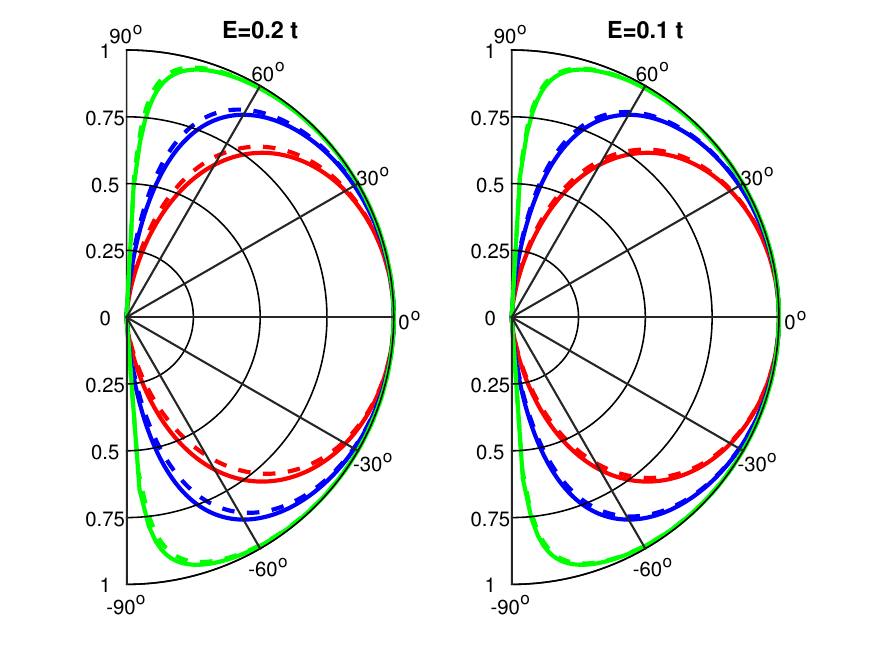}
      \includegraphics[width=0.5\textwidth]{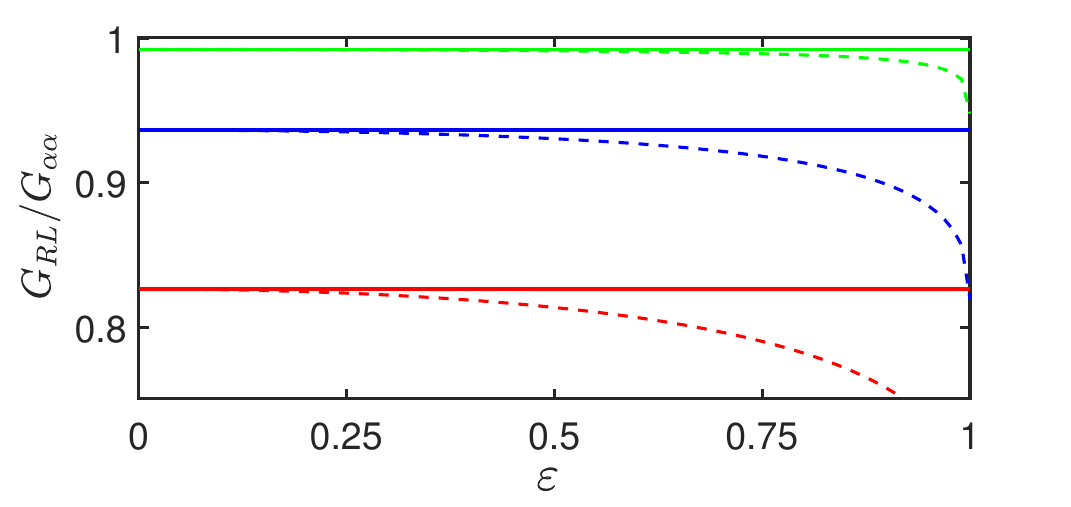}  
\caption{Top panel: Transmission probabilities $T_{\textrm{cont}}(\theta)$ and $T_{\textrm{latt}}(\varepsilon,\theta)$ through the straight domain wall in terms of the incidence angle $\theta$ and for different values of the energy $\varepsilon$. 
Bottom panel: Conductance (in units of the conductance $G_{\alpha\alpha}$ when $\alpha_R=\alpha_L$ ) as a function of the energy. Solid lines for the low-energy continuum description and dashed lines for the lattice description.
Green line $\Delta_{RL}=0.055$, blue line $\Delta_{RL}=0.22$ and red line $\Delta_{RL}=0.5$.}
\label{figure1}
\end{figure}

As illustrated in Fig. \ref{fig_contour},
for a given energy $\varepsilon>0$ and a wave vector $(k_x,k_y)$ around the Dirac point $K=(0,\frac{4\pi}{3\sqrt{3}})$, we can write
$(k_x,k_y)=K+q(\cos \theta,\sin \theta)$, which implicitly define the effective incident angle $\theta$, where the modulus $q$   generally depends on both $\varepsilon$ and $\theta$. More explicitly, the effective incidence angle
writes
\begin{equation}
\theta=\arctan \frac{k_y-\frac{4\pi}{3\sqrt{3}}}{k_x}.
\label{ky teta1}
\end{equation}
This allows us to parametrically express the transmission probability $T_{\textrm{latt}}(\varepsilon,\theta)$ as a function of the energy $\varepsilon$ and the incidence angle $\theta$. A direct comparison  between the lattice model result $T_{\textrm{latt}}(\varepsilon,\theta)$ and the continuum  model prediction $T_{\textrm{cont}}(\theta)$ is shown in  Fig. \ref{figure1} for various energy $\varepsilon$ and different values of the Berry phase jump $\Delta_{RL}$. 
 Moreover, the conductance obtained from the continuum description [see Eq. (\ref{cond_cont})] is displayed in Fig. \ref{figure1} and compared with the conductance calculated from the lattice description.
As expected for sufficiently low-energy $\varepsilon$, corresponding to circular energy contour, there is excellent quantitative agreement between the lattice and continuous model results. Indeed, around the Dirac point $K(0,\frac{4\pi}{3\sqrt{3}})$ the expression of the parameter $\Gamma(\varepsilon \sim 0,k_y=q_y+\frac{4\pi}{3\sqrt{3}})$ [Eq. (\ref{Gamma})], which appears in the expression of the transmission probability $T_{\textrm{latt}}$ [Eq. (\ref{T type1})], tends toward $\frac{q_y}{q_x}=\tan \theta$. 
Consequently, in the low-energy limit, the lattice transmission converges to the continuum result $T_{\textrm{latt}}(\varepsilon \sim 0,\theta) \rightarrow T_{\textrm{cont}}(\theta) $.
For larger values of $\varepsilon$, the lattice model results show an asymmetry $\theta \rightarrow -\theta$, reflecting the trigonal warping of the energy contour,  as shown in Fig. \ref{fig_contour}.

\subsection{Scattering properties of the zigzag domain wall}

We now detail the scattering properties of the lattice model for the
 zigzag domain wall, composed of $A$ and $C$ sites located at positions
${\bm r}_A+m \bm{a}_2$ and  ${\bm r}_C+m \bm{a}_2$ (and $n=0$), as depicted in Fig. \ref{figtype2}.

For this type of domain wall, the key change concerns the $n=0$ interface condition that now becomes
\begin{equation}
\left\{
 \begin{array}{l}
  \varepsilon \psi_{0}^A=c_{R} \psi_{0}^B+c_{L} (1+ z_2) \psi_{-1}^B,\\
  \varepsilon \psi_{0}^C=s_{R}(1+z_2)\psi_{0}^B+s_{L}  z_2\psi_{-1}^B,\\
  \begin{split}
  \varepsilon \psi_{0}^B=&c_{R}[ \psi_{0}^A +(1+z_2^*)\psi_{1}^A]+\\&s_{R}[ (1+z_2^*)\psi_{0}^C+z_2^*\psi_{1}^C].
  \end{split}
 \end{array}\right.
\end{equation}
As a consequence, we obtain the following modified set of equations
\begin{equation}
\begin{array}{ll}
 n<0 &  |\Psi_{n}\rangle= M_{L} |\Psi_{n-1}\rangle,\\
   & |\Psi_{0}\rangle= M_{RL} |\Psi_{-1}\rangle,\\
 n>0 &  |\Psi_{n}\rangle= M_{R} |\Psi_{n-1}\rangle,
 \end{array}
 \label{transfer1}
\end{equation}
with a modified interface transfer matrix given by
\begin{equation}
\begin{split}
M_{RL}(\varepsilon,k_y)=
&\left(
 \begin{array}{ccc}
        \varepsilon &-c_{R} & 0\\
        -c_{L}(1+z_2^*) &0&- s_{L} z_2^*  \\
          0&-s_{R} (1+z_2)  &\varepsilon
\end{array}
\right)^{-1}\\& \times
 \left(
     \begin{array}{ccc}
        0 &c_{L} (1+z_2)& 0 \\
        c_{L} &-\varepsilon& s_{L}(1+z_2^*)\\
        0& s_{L} z_2  & 0
       \end{array}
       \right),
       \end{split}
       \label{trans mat typ1}
\end{equation}
From there, we can now consider a generic {\em itinerant state} in the $L$ region.
It now arrives on the left of the interface ($n=-1$) with a form
 $|\Psi_{-1}\rangle= a_+ \lambda_+^{-1}|\Psi^{L}_{+}\rangle +a_- \lambda_-^{-1}|\Psi^{L}_{-}\rangle$ with necessarily $a_0=0$. After the interface it is scattered to a state ($n=0$)
 $|\Psi_{0}\rangle=M_{R L} |\Psi_{-1}\rangle=b_+ |\Psi^{R}_{+}\rangle +b_- |\Psi^{R}_{-}\rangle+b_0 |\Psi^{R}_{0}\rangle$
 with $b_-=0$ (same reasoning as in the previous case). Therefore, we can write
\begin{equation}
{\cal N}_{s'} b_{s'}=\sum_{s=\pm}{M}_{s's} a_s \lambda_s^{-1},\\
\end{equation}
with
\begin{equation}
 {M}_{s's}=\llangle \Psi^{R}_{s'}|M_{R L}|\Psi^{L}_s\rangle.
\end{equation}
More quantitatively, we obtain
\begin{equation}
 \begin{array}{l}
M_{\pm\pm}=(1+z_2)\dfrac{c^2_{R L}\lambda_\mp-\lambda_\pm}{c_{R L}}\lambda_\pm^2,\\
M_{\pm \mp}=-z_2(1+z_2)\dfrac{s^2_{R L}}{c_{R L}}\lambda_\mp,\\
M_{0\pm}=z_2(1+z_2)s_{R L}\lambda_\pm.
 \end{array}
\end{equation}
with $c_{R L}$ and $s_{R L}$ are given in Eq. (\ref{crl_srl}). 

Writing $a_-=r a_+$, $b_+=r a_+$ and $b_0=t_0 a_+$, we obtain
\begin{equation}
 \begin{array}{l}
r=-\frac{{M}_{-+}\lambda_-}{{M}_{--}\lambda_+}
=\dfrac{s^2_{R L} }{c^2_{R L}-z_{\pm}},\\
t=\frac{{M}_{++}{M}_{--}-{M}_{-+}{M}_{+-}}{{\cal N}_{+}{M}_{--}\lambda_+}
=\dfrac{ c_{R L}(1-z_{\pm})}{ c^2_{R L}-z_{\pm}},\\
t_0=\frac{{M}_{0+}{M}_{--}-{M}_{0-}{M}_{-+}}{{\cal N}_{0}{M}_{--}\lambda_+}
=\dfrac{s_{R L}(1-z_{\pm})}{c^2_{R L}-z_{\pm}},\\
 \end{array}
 \label{t type2}
\end{equation}
where $z_{\pm}=\lambda_-/\lambda_+$.
The transmission probability is given by 
\begin{equation}
T_{\textrm{latt}}(\varepsilon,\theta)=|t|^2=\dfrac{\sin^2(\frac{3}{2}k_x)}{\sin^2(\frac{3}{2}k_x)+\frac{s_{R L}^4}{4c^2_{R L}}},
\label{trans_zigzag_latt}
\end{equation}
where $k_x$ is a function of the energy $\varepsilon$ and the incidence angle $\theta$ as in the previous section [see Eqs. (\ref{kx_e_ky}) and (\ref{ky teta1})].
In the low-energy limit, writing $\frac{3}{2}k_x  \propto \varepsilon \cos \theta$ and defining the effective energy scale
$\varepsilon_{R L}=\frac{s_{R L}^2}{2c_{R L}}$,
we obtain the approximate expression
\begin{equation}
 T_{\textrm{latt}}(\varepsilon,\theta)\simeq \frac{\varepsilon ^2 \cos^2 \theta}{\varepsilon ^2 \cos^2 \theta+\varepsilon_{R L}^2}.
 \label{transmi2}
\end{equation}
The main striking feature is that $T_{\textrm{latt}}(\varepsilon,\theta)$ vanishes at $\varepsilon=0$. Moreover, there is no Klein tunneling effect at normal incidence $\theta=0$. Nevertheless, $\theta=0$ corresponds to the maximum of the transmission probability for any $\varepsilon$, with $T_{\textrm{latt}}(\varepsilon,\theta=0)\le 1/2$ for $\varepsilon \le \varepsilon_{R L}$ and $T_{\textrm{latt}}(\varepsilon,\theta=0)\ge 1/2$ for $\varepsilon \ge \varepsilon_{R L}$.

Another interesting feature of the zigzag domain wall is that there is a finite probability $T_0$ of propagating along the domain wall.
Accounting for the different normalizations of the scattering states $|\Psi_{0}^{\alpha}\rangle$ and $|\Psi_{+}^\alpha\rangle$,
this probability can be expressed as $T_0=\frac{N_0}{N_+}|t_0|^2$, which explicitly reads as follows
\begin{equation}
 T_0(k_x,k_y)=\frac{2+2\cos{\sqrt{3}k_y}s^2_{R}-c^2_{R}}{2\varepsilon^2}\frac{s_{R L}^2}{c_{R L}^2}  T_{\textrm{latt}}(k_x,k_y)
\end{equation}

We have no clear understanding of the physical mechanism at the origin of this possibility to propagate along
the domain-wall interface.

\begin{figure}[h!]
       \centering
     \includegraphics[width=0.4\textwidth]{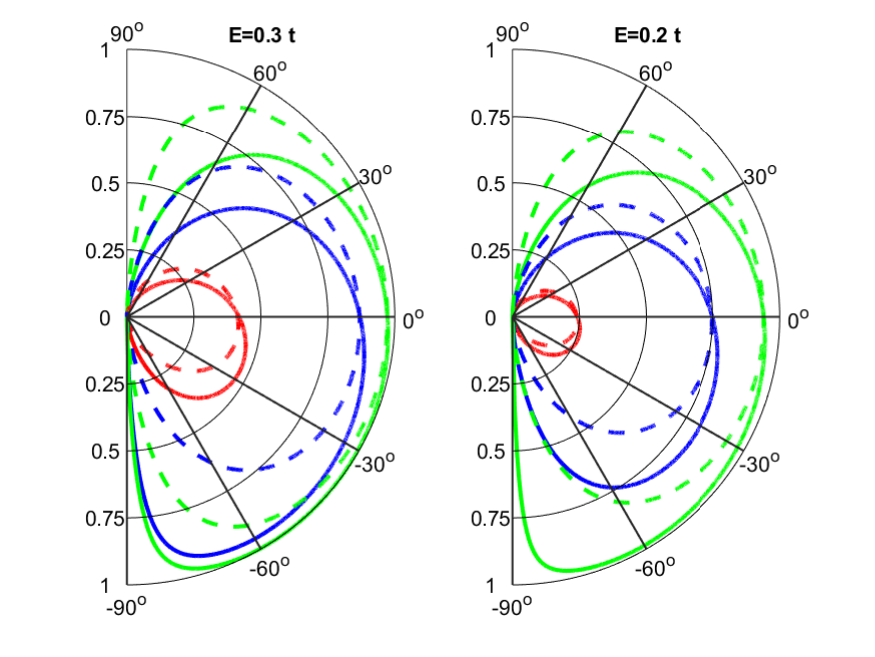}
     \includegraphics[width=0.4\textwidth]{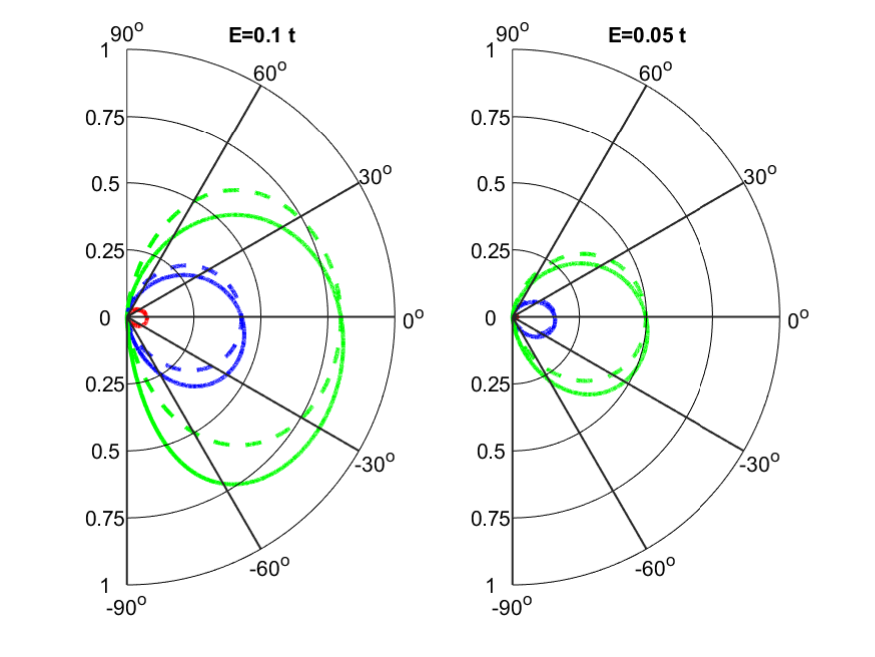}
     \includegraphics[width=0.5\textwidth]{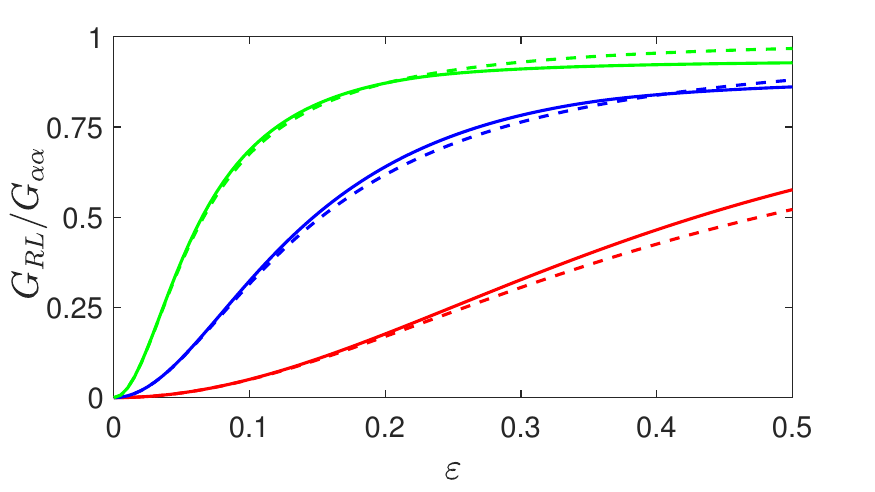}

\caption{ 
Top panel: Transmission probabilities $T_{\textrm{cont}}(\theta,\varepsilon)$ and $T_{\textrm{latt}}(\varepsilon,\theta)$  through a zigzag domain wall as functions of the incidence angle $\theta$, shown for different energy values $\varepsilon$. 
Bottom panel: Conductance (in units of the conductance $G_{\alpha\alpha}$ when $\alpha_R=\alpha_L$ ) as a function of energy. Solid lines correspond to the low-energy continuum model, while dashed lines represent the lattice model.
Red line $\varepsilon_{RL}=0.35$, $\Delta_{RL}=0.5$ , blue line $\varepsilon_{RL}=0.11$, $\Delta_{RL}=0.32$ and green line $\varepsilon_{RL}=0.05$, $\Delta_{RL}=0.22$.}
\label{figure7v}
\end{figure}

We note that the low-energy limit of the transmission probability in Eq. (\ref{transmi2}) can be recovered using the continuum approach, by employing an effective matching matrix of the general form given in Eq. (\ref{genematch matr}) with suitable values for the parameters $\lambda_i$.
More explicitly, as shown in section II, by taking $\lambda_1=1$, $\lambda_2=0$ 
and choosing $\lambda_3=2\varepsilon_{RL}/\varepsilon$, the transmission probability, given by Eq. (\ref{cont type4}) becomes
\begin{equation}
T_{\textrm{cont}}(\theta,\varepsilon)=\frac{\cos^2 \theta}{\cos^2 \theta +(\Delta_{RL}\sin\theta+\varepsilon_{RL}/\varepsilon)^2}.
\label{trans_cont_zigzag}
\end{equation}
While the expressions $T_{\textrm{cont}}(\theta,\varepsilon)$ and $T_{\textrm{latt}}(\varepsilon, \theta)$ [ Eq. (\ref{transmi2})] are different at low-energy, they show good quantitative agreement, as illustrated in Fig. \ref{figure7v} for various energy values $\varepsilon$ and different effective energy scales, defined as $\varepsilon_{R L}=\frac{s_{R L}^2}{2c_{R L}}$, as well as different Berry phase jumps $\Delta_{RL}$.

Finally, as in the previous case, the corresponding ballistic conductance [Eq. (\ref{conductance1})] can be expressed analytically as
\begin{equation}
G_{RL}=G_{\alpha\alpha} \left[ 1+(1/a_{RL}-a_{RL})\operatorname{artanh}(1/a_{RL})\right]
\label{cond_cont_zg}
\end{equation}
where $G_{\alpha\alpha}=\frac{4e^2}{\pi h}\frac{L_y E}{\hbar v_F}$ is the conductance in the case of perfect transmission  (i.e., when $\alpha_R=\alpha_L$), $a_{RL}=[1+(\varepsilon_{RL}/\varepsilon)^2]^{1/2}$. Here, $L_y$ denotes the sample width in the $y$ direction and the factor $4$ accounts for the spin and valley degeneracy.

A direct comparison between the lattice model transmission probability, $T_{\textrm{latt}}(\varepsilon, \theta)$ [Eq. (\ref{trans_zigzag_latt})], and the continuum model result, $T_{\textrm{cont}}(\theta)$ [Eq. (\ref{trans_cont_zigzag})], is presented in Fig. \ref{figure7v} for various energy values $\varepsilon$ and different effective energy scales, defined as $\varepsilon_{R L}=\frac{s_{R L}^2}{2c_{R L}}$. In addition, the conductance calculated within the continuum model (see Eq. (\ref{cond_cont_zg})) is also shown in Fig. \ref{figure7v} and compared with the conductance obtained from the lattice model.

\section{Conclusion}
In this work, we have investigated the scattering properties of a 2D massless Dirac particle through a domain wall that separates two regions with distinct quantum geometry (distinct Berry phase). More concretely, we have used the $\alpha-T_3$ model with different parameters $\alpha_L$ and $\alpha_R$ on each side of the domain wall such that there is a (nonquantized) Berry phase jump at the domain-wall interface. Importantly, for such an interface the transmitted (refracted) and incident momenta are equal.

In a first step, we have used a low-energy continuum description (valid near the Dirac point) in which the parameter $\alpha$ determines the nature of the effective pseudospin of the Dirac particle on each side of the domain wall. We have shown that there is a partial transmission with a probability that solely depends on the incident angle and on the Berry phase jump at the domain-wall interface but does not depend on the particle’s energy.
This result already contrasts with usual potential barriers for which the transmission (reflection) probability is directly proportional to the difference between the incident and transmitted (refracted) momenta, as well as on the energy of the incident particle. However, similarly to usual potential barrier, at zero incidence angle we obtain perfect transmission (Klein tunneling) whatever the Berry phase jump.

In a second step, we have examined the scattering properties across an $\alpha_L-\alpha_R$ domain wall within the lattice tight-binding $\alpha-T_3$ model.
In that situation, two distinct types of domain walls must be distinguished.
For the two kinds of domain walls the transmission probability depends both on the Berry phase jump at the interface, the incident angle and the energy of the Dirac particle.

For the straight domain wall, as presented in the main text, the lattice transmission probability shows a Klein tunneling effect at all energies.
 At sufficiently low energies, it also shows excellent quantitative agreement with the continuum model for all angles of incidence.
For larger energies, the lattice transmission probability shows an asymmetry $\theta \rightarrow -\theta$ with the incident angle $\theta$. 
This asymmetry arises from trigonal warping, which distorts the constant energy contours away from a perfect circle.

For the zigzag domain wall, the effective transmission probability does not show the Klein tunneling effect and varies more strongly with the energy and the Berry phase jump. More quantitatively, the smaller the energy of the incident particle, the smaller the transmission probability 
and the more pronounced the effect of the Berry phase jump.
For this zigzag domain wall, the lattice model reveals that, in addition to the reflected and transmitted waves, there is a finite probability for the particle to propagate along the domain wall.
Furthermore, we show that, in the low-energy limit, the transmission probability for this domain wall can be recovered using a continuum model with a more general form of the interface matching matrix.
In the appendix, we further demonstrate that such an effective matching matrix can be derived as the result of an interface potential within the continuum model. 

On a larger perspective, our results indicate that spatial inhomogeneities or defects in quantum geometry may serve as alternative mechanisms for the scattering of Dirac particles. We believe that it would be interesting to consider the scattering of Dirac particles in the $\alpha-T_3$ continuum and lattice models with different kinds of spatial variation of the $\alpha$ parameter; for example a circular domain wall or a spatial {\em vortex} structure of the $\alpha$ parameter.

We believe that our predictions can be experimentally tested using a two-dimensional phononic crystal with a triangular lattice, as proposed in Ref. \cite{Zhu2023}. In our setup, the domain wall separates a pseudospin-1 phononic crystal from a pseudospin-1/2 phononic crystal. This configuration is equivalent to a domain wall characterized by $\alpha_L=1$ and $\alpha_R=0$. 
Note that the carriers in the phononic crystal are bosons; this does not affect our results, since the transmission probability is independent of quantum statistics, unlike the conductance.

\label{sectionIV}
\section*{Acknowledgments}
This work was supported by the Tunisian Ministry of
Higher Education and Scientific Research. L.M. acknowledges
the LPS in Orsay for financial support and kind hospitality, where the work started.

\section*{Appendix: Matching matrix in the continuum model with an interface potential}
In this appendix, we explain how the general matching matrix [Eq. (\ref{genematch matr})] given in the main text may be viewed as resulting from the continuum model with an additional interface potential.

\subsection{Domain wall without an interface potential}

We first start by considering the case without an interface potential.
The effective one-dimensional continuum Hamiltonian model may be written
\begin{equation}
 H(x)=
\left\lbrace\begin{array}{ll}
H_{L} & x<0,\\
H_{R} & x>0.
     \end{array}\right.
\end{equation}
with
\begin{equation}
 H_\alpha =\hbar v_F(-i\partial_x S_y^\alpha +\xi q_y S_x^\alpha),
\end{equation}
for $\alpha=L,R$.
Without loss of generality, we can formally write the scattering state of energy $\varepsilon$ as
$\Psi(x)=\Psi_L(x) \theta(-x)+ \Psi_R(x) \theta(x)$.
The main difficulty with the above Hamiltonian arises from the fact
that the current operator along the $x$ direction is
different on both sides of the interface. As a consequence,
the scattering states verify an effective interface condition of the form
\begin{equation}
S^R_y \Psi_R(0^+)={\cal M}_{RL} S^L_y \Psi_L(0^-),
\end{equation}
with ${\cal M}_{RL}$ a nontrivial $3\times 3$ matching matrix.
To obtain the matching matrix ${\cal M}_{RL}$,
a convenient way consists of {\it rotating} each $H_\alpha$
to the basis that diagonalizes $S_\alpha ^y$.
The corresponding orthogonal transformation is given by
\begin{equation}
 O_{\alpha}=\frac{1}{\sqrt{2}}\begin{pmatrix}
c_{\alpha} & \sqrt{2}s_{\alpha}  & c_{\alpha}\\
i & 0 & -i\\
-s_{\alpha} & \sqrt{2}c_{\alpha}  & -s_{\alpha}
\end{pmatrix},
\label{o_alpha}
\end{equation}
such that ${\bar H}_\alpha=O_{\alpha}^{\dag}  H_\alpha O_{\alpha}$ rewrites
\begin{equation}
 {\bar H}_\alpha=\hbar v_F(-i\partial_x S_z +\xi q_y{\bar S_x}^\alpha),
\end{equation}
with $S_z=O_{\alpha}^{\dag}  S_y ^\alpha O_{\alpha}$ and ${\bar S_x}^\alpha=O_{\alpha}^{\dag}  S_x ^\alpha O_{\alpha}$.
The effective Hamiltonian in the rotated bases may now be written
\begin{equation}
{\bar H}(x)=\hbar v_F(-i\partial_x S_z +\xi q_y {\bar S_x}(x)),
\end{equation}
where ${\bar S_x}(x)={\bar S_x}^L \theta(-x)+{\bar S_x}^R\theta(x)$.
Writing now $\Phi(x)=\Phi_L(x) \theta(-x)+ \Phi_R(x) \theta(x)$ the scattering states in the rotated bases with $\Psi_\alpha(x)=O_{\alpha} \Phi_\alpha(x)$.
Since the current operator is the same on both sides of the interface in the rotated basis,
 we can now proceed in the usual manner in order to obtain the interface boundary condition. More concretely, applying the operator $-i\partial_x S_z$
 to the scattering state $\Phi(x)$ one obtains two kinds of contributions: \\
 (i) $ i\hbar v_F(\theta(-x)  \partial_x \Phi_L(x)+ \theta(x) \partial_x \Phi_R(x))$, and \\
   (ii) $ i\hbar v_F S_z (-\delta(-x) \Phi_L(x)+ \delta(x) \Phi_R(x))$.\\
   When writing $H \Phi(x)=\varepsilon \Phi(x)$, the contributions in (i) lead to the usual eigenvalue equations for the scattering states in the regions $x<0$ and $x>0$.
The contributions in (ii), which contain $\delta$ functions at the interface, yield the boundary condition
\begin{equation}
 S_z\Phi_R(0)=S_z\Phi_L(0).
\end{equation}
Going back to the original nonrotated bases, we deduce that the effective matching matrix is given by
${\cal M}_{RL}=O_{R}O_{L}^{\dagger}$ as presented in the main text.

\subsection{Domain wall with an interface potential}

We now consider the effect of an effective interface potential in the rotated bases. 
Taking a generic form ${\bar V}(x)=\delta(x) {\bar V}$ with ${\bar V}$ a $3 \times 3$ matrix,
the effective interface condition becomes
\begin{equation}
 (i \hbar v_F S_z-{\bar V}/2) \Phi_R(0)=(i \hbar v_F S_z+{\bar V}/2) \Phi_L(0),
\end{equation}
where we have taken the convention $\theta(0)=1/2$ for the Heaviside function.
Let's assume that the matrix $V$ may be written ${\bar V}=i \hbar v_F A S_z$
with $A$ a $3 \times 3$ matrix. To leading order in $V$, the interface condition now reads
\begin{equation}
  S_z \Phi_R(0)=(1+A)S_z \Phi_L(0).
\end{equation}
Transforming back to the original, nonrotated basis, the corresponding interface condition takes the form
$S^R_y \Psi_R(0^+)={\cal M}_{RL} S^L_y \Psi_L(0^-)$ with an effective matching matrix given by
\begin{equation}
 {\cal M}_{RL}=O_{R}(1+A)O_{L}^{\dagger}.
\end{equation}
To obtain an effective matching matrix ${\cal M}_{RL}$ with the general form [Eq.(\ref{genematch matr})]
in the main text, the corresponding matrix $A$ is obtained as
\begin{equation}
A= (O_{R}^{\dagger}M_{RL}O_{L}-1)
\end{equation}
such that
the effective interface potential reads
\begin{equation}
{\bar V}=i \hbar v_F A S_z=i \hbar v_F
\begin{pmatrix}
a&0&b\\
0&0&0\\
-b^*&0&-a^*
\end{pmatrix},
\end{equation}
with
\begin{equation}
 \begin{array}{l}
  a=-1+\frac{1}{2}(\lambda_1+\lambda_4+ i(\lambda_2+\lambda_3))\\
  b=\frac{1}{2}(\lambda_1-\lambda_4+ i(\lambda_2-\lambda_3)).
 \end{array}
\end{equation}
Note that up to this point, we have mainly established that the general matching matrix ${\cal M}_{RL}$
can be interpreted as arising from an effective interface potential ${\bar V}(x)$ in the rotated basis.
The last necessary step 
consists of determining the corresponding interface potential in the original (nonrotated) basis and comparing it with the continuum limit of the effective lattice interface potential given by Eq. (\ref{interfacepotential2}) in the main text.

\end{document}